\documentstyle[preprint,aps,epsf,floats,eqsecnum]{revtex}
\begin{document}
\tighten

\def\bfl{{\bbox \ell}}
\newcommand{\gsim}{\raisebox{-0.7ex}{$\stackrel{\textstyle >}{\sim}$ }}
\newcommand{\lsim}{\raisebox{-0.7ex}{$\stackrel{\textstyle <}{\sim}$ }}
\def\pislash{ {\pi\hskip-0.6em /} }
\def\pislashsmall{ {\pi\hskip-0.375em /} }
\def\pslash{p\hskip-0.45em /}
\def\nopi{ {\rm EFT}(\pislash) }
\def\Zpi{ {^\pislashsmall \hskip-0.25em Z} }
\def\bull{\vrule height .9ex width .8ex depth -.1ex}
\def\MeV{{\rm MeV}}
\def\GeV{{\rm GeV}}
\def\Tr{{\rm Tr\,}}
\def\nrcpt{NR\raise.4ex\hbox{$\chi$}PT\ }
\def\ket#1{\vert#1\rangle}
\def\bra#1{\langle#1\vert}
\def\ltap{\ \raise.3ex\hbox{$<$\kern-.75em\lower1ex\hbox{$\sim$}}\ }
\def\gtap{\ \raise.3ex\hbox{$>$\kern-.75em\lower1ex\hbox{$\sim$}}\ }
\def\abs#1{\left| #1\right|}
\def\CA{{\cal A}}
\def\CC{{\cal C}}
\def\CD{{\cal D}}
\def\CE{{\cal E}}
\def\CL{{\cal L}}
\def\CO{{\cal O}}
\def\CZ{{\cal Z}}
\def\bvert{\Bigl\vert\Bigr.}
\def\pds{{\it PDS}\ }
\def\ms{MS}
\def\ddq{{{\rm d}^dq \over (2\pi)^d}\,}
\def\ddqm{{{\rm d}^{d-1}{\bf q} \over (2\pi)^{d-1}}\,}
\def\bfq{{\bf q}}
\def\bfk{{\bf k}}
\def\bfp{{\bf p}}
\def\bfpp{{\bf p '}}
\def\bfr{{\bf r}}
\def\dtr{{\rm d}^3\bfr\,}
\def\bfx{{\bf x}}
\def\dtx{{\rm d}^3\bfx\,}
\def\dfx{{\rm d}^4 x\,}
\def\bfy{{\bf y}}
\def\dty{{\rm d}^3\bfy\,}
\def\dfy{{\rm d}^4 y\,}
\def\dfq{{{\rm d}^4 q\over (2\pi)^4}\,}
\def\dfk{{{\rm d}^4 k\over (2\pi)^4}\,}
\def\dfl{{{\rm d}^4 \ell\over (2\pi)^4}\,}
\def\dtq{{{\rm d}^3 {\bf q}\over (2\pi)^3}\,}
\def\dtk{{{\rm d}^3 {\bf k}\over (2\pi)^3}\,}
\def\dtl{{{\rm d}^3 {\bfl}\over (2\pi)^3}\,}
\def\dt{{\rm d}t\,}
\def\frac#1#2{{\textstyle{#1\over#2}}}
\def\darr#1{\raise1.5ex\hbox{$\leftrightarrow$}\mkern-16.5mu #1}
\def\){\right)}
\def\({\left( }
\def\]{\right] }
\def\[{\left[ }
\def\si{{}^1\kern-.14em S_0}
\def\siii{{}^3\kern-.14em S_1}
\def\diii{{}^3\kern-.14em D_1}
\def\fm{{\rm\ fm}}
\def\MeV{{\rm\ MeV}}
\def\CA{{\cal A}}

\def\Czero{ {^\pislashsmall \hskip -0.2em C_0^{(\siii)} }}
\def\Ctwo{ {^\pislashsmall \hskip -0.2em C_2^{(\siii)} }}
\def\Cfourone{ {^\pislashsmall \hskip -0.2em \tilde C_4^{(\siii)} }}
\def\Cfourtwo{ {^\pislashsmall \hskip -0.2em C_4^{(\siii)} }}
\def\Cn{ {^\pislashsmall \hskip -0.3em C_{2n} }}
\def\Czerominus{ {^\pislashsmall \hskip -0.2em C_{0,-1}^{(\siii)} }}
\def\Czerominussing{ {^\pislashsmall \hskip -0.2em C_{0,-1}^{(\si)} }}
\def\Czerozero{ {^\pislashsmall \hskip -0.2em C_{0,0}^{(\siii)} }}
\def\Czeroone{ {^\pislashsmall \hskip -0.2em C_{0,1}^{(\siii)} }}
\def\Ctwozero{ {^\pislashsmall \hskip -0.2em C_{2,-2}^{(\siii)} }}
\def\Ctwozerosing{ {^\pislashsmall \hskip -0.2em C_{2,-2}^{(\si)} }}
\def\Ctwoone{ {^\pislashsmall \hskip -0.2em C_{2,-1}^{(\siii)} }}
\def\Cfourtwoone{ {^\pislashsmall \hskip -0.2em C_{4,-3}^{(\siii)} }}
\def\CSDzero{ {^\pislashsmall \hskip -0.2em C_0^{(sd)} }}
\def\CSDtwoone{ {^\pislashsmall \hskip -0.2em \tilde C_2^{(sd)} }}
\def\CSDtwotwo{ {^\pislashsmall \hskip -0.2em C_2^{(sd)} }}
\def\CSDtwotwotwo{ {^\pislashsmall \hskip -0.2em C_{2,-2}^{(sd)} }}
\def\CSDzeroone{ {^\pislashsmall \hskip -0.2em C_{0,-1}^{(sd)} }}
\def\CSDzerotwo{ {^\pislashsmall \hskip -0.2em C_{0,0}^{(sd)} }}
\def\Ltwo{ {^\pislashsmall \hskip -0.2em L_2 }}
\def\Lone{ {^\pislashsmall \hskip -0.2em L_1 }}
\def\CQuad{ {^\pislashsmall \hskip -0.2em C_{\cal Q} }}

\def\Ames{ A }  

\newcommand{\eqn}[1]{\label{eq:#1}}
\newcommand{\refeq}[1]{(\ref{eq:#1})}
\newcommand{\eq}{eq.~\refeq}
\newcommand{\eqs}[2]{eqs.~(\ref{eq:#1}-\ref{eq:#2})}
\newcommand{\eqsii}[2]{eqs.~(\ref{eq:#1}, \ref{eq:#2})}
\newcommand{\Eq}{Eq.~\refeq}
\newcommand{\Eqs}{Eqs.~\refeq}

\def\Journal#1#2#3#4{{#1} {\bf #2}, #3 (#4)}

\def\NCA{\em Nuovo Cimento}
\def\NIM{\em Nucl. Instrum. Methods}
\def\NIMA{{\em Nucl. Instrum. Methods} A}
\def\NPB{{\em Nucl. Phys.} B}
\def\NPA{{\em Nucl. Phys.} A}
\def\NP{{\em Nucl. Phys.} }
\def\PLB{{\em Phys. Lett.} B}
\def\PRL{\em Phys. Rev. Lett.}
\def\PRD{{\em Phys. Rev.} D}
\def\PRC{{\em Phys. Rev.} C}
\def\PRA{{\em Phys. Rev.} A}
\def\PR{{\em Phys. Rev.} }
\def\ZPC{{\em Z. Phys.} C}
\def\SJP{{\em Sov. Phys. JETP}}
\def\SJNP{{\em Sov. Phys. Nucl. Phys.}}

\def\FBS{{\em Few Body Systems Suppl.}}
\def\IJMP{{\em Int. J. Mod. Phys.} A}
\def\UJP{{\em Ukr. J. of Phys.}}
\def\CJP{{\em Can. J. Phys.}}
\def\SCI{{\em Science} }


\preprint{\vbox{
\hbox{ NT@UW-99-14}
}}
\bigskip
\bigskip

\title{Nucleon-Nucleon Effective Field Theory Without Pions}
\author{Jiunn-Wei Chen$^a$, Gautam Rupak$^a$ and Martin  J. Savage$^{a,b}$}  
\address{ $^a$ Department of Physics, University of Washington,  
Seattle, WA 98915. }
\address{ $^b$ Jefferson Lab., 12000 Jefferson Avenue, Newport News, 
Virginia 23606.}
\maketitle

\begin{abstract}
Nuclear processes involving momenta much below the mass of the pion may be
described by an effective field theory in which the pions do not appear as
explicit degrees of freedom.  The effects of the pion and all other virtual
hadrons are reproduced by the coefficients of gauge-invariant
local operators involving the nucleon field.
Nucleon-nucleon scattering phase shift data constrains many of the
coefficients that appear in the effective Lagrangean
but at
some order in the expansion coefficients enter that must be constrained by
other observables.
We compute several observables in the two-nucleon sector up to next-to-next-to
leading order in the effective field theory without pions,
or to the order at which
a counterterm involving four-nucleon field operators is encountered.
Effective range theory is recovered from the effective field theory
up to the order
where relativistic corrections enter or where four-nucleon-external current
local operators arise.
For the deuteron magnetic moment, quadrupole moment
and the $np\rightarrow d\gamma$ radiative capture cross section
a four-nucleon-one-photon counterterm exists at next-to-leading order.
The electric polarizability and electric charge form factor of the
deuteron are determined 
up to next-to-next-to-leading order, which includes the first
appearance of relativistic corrections.
\end{abstract}

\vskip 2in

\leftline{February 1999}
%
%
%
%
\vfill\eject


\section{Introduction}

Weinberg's  pioneering work\cite{Weinberg1}
on the application of effective field theory (EFT) 
to nuclear physics has given rise to a decade long effort to construct a
perturbative theory of nuclear interactions\cite{Bira}-\cite{threebod}.
The small kinetic and potential energies of nuclear systems
compared to
the typical scale of strong interactions, either $\Lambda_{\rm QCD}$ or the
chiral symmetry breaking scale $\Lambda_\chi$, suggests that the ratio of these
two
scales might provide a small expansion parameter.
Indeed, in the single nucleon sector the ratio of external momenta and quark
masses, $m_q$,  to $\Lambda_\chi$ allows for a systematic field theory
description of pionic and external current interactions with a single nucleon.
However, extending this simple idea to multi-nucleon systems has proved to be
a challenge.

It is important to understand nuclear physics directly from quantum
chromodynamics (QCD) and constructing the
correct low-energy effective field theory is a central component of such a
program.  The
optimal way to determine the properties of multi-nucleon systems is to
match onto the low-energy effective field theory and perform computations
directly with the effective theory,
and not with the theory
written in terms of quark and gluon fields.
Further, one wants to be able to make such a matching between QCD and the
low-energy effective theory free of any unjustified
assumptions about the dynamics or structure of the hadrons.
Unfortunately, at this point in time such a matching is not possible.
However, the symmetries of QCD along with the fact that we wish to examine
observables involving small external momenta allow the low-energy
effective field theory to relate the amplitudes of different strong interaction
processes, at some level of precision.

Significant progress in understanding both the two-body
and three-body systems in terms of effective field theory 
has been made during the last year.
Several observables in the two-nucleon sector have been explored in detail in
the theory with dynamical pions, such as the electromagnetic form factors of
the deuteron~\cite{KSW2}, $\gamma d\rightarrow\gamma d$~\cite{CGSSpol,Ccompt},
$np\rightarrow d\gamma$~\cite{KSSWpv,SSWst,Parkeft} and $\nu d\rightarrow \nu d
, \nu n p $\cite{BCnu}.
In addition, the inclusion of coulomb interactions into the pionless
theory has been shown to be straightforward\cite{KRpp}.
In the three-nucleon system it has been demonstrated
that 
low-momentum processes can be accurately and precisely described
in the effective field theory without dynamical pions~\cite{threebod}.
For example the quartet $N-d$ scattering length has a rapidly converging
expansion in terms of the ratio $r_0^{(\siii)}/a^{(\siii)}$,
where $r_0^{(\siii)}$ is the effective range and
$a^{(\siii)}$ is the scattering length for $NN$ scattering in the
spin-triplet channel.
Calculations in the two-nucleon 
sector\cite{KSW2,CGSSpol,Ccompt,SSpv,KSSWpv,SSWst}
with dynamical pions
using the KSW power counting scheme~\cite{KSW},
(where pions and higher dimension four-nucleon operators are treated in
perturbation theory)  have been performed at leading order (LO) and 
next-to-leading order (NLO), and just 
recently been extended in some cases to 
next-to-next-to-leading (NNLO)~\cite{MehStew,GruSho}.

In this work we make a thorough investigation of the two-nucleon sector
in the theory without dynamical pions, appropriate for very low-momentum
processes.
Effective range theory\cite{ERtheory} is found to 
reproduce the effective field theory
up to the order where relativistic corrections or
local four-nucleon-external-current local operators enter.
In many cases effective field theories for hadronic systems 
are used to systematically include
the approximate $SU(2)_L\otimes SU(2)_R$ chiral symmetry of QCD.
As we are considering the theory of nucleon interactions
where pions do not explicitly appear, chiral
symmetry does not play a central role and the usefulness of this investigation
could be questioned.
One motivation for constructing and understanding this theory is that
calculations
of the observables considered in this work
with any scheme, such as potential models or the effective field theory with
dynamical pions,
must reproduce the analytic results we obtain.
A secondary motivation is that this theory provides a very clear demonstration
of how the perturbative effective field theory approach to $NN$ interactions
behaves at higher orders, including relativistic effects.


\section{NN Scattering in the $\siii-\diii$ Channel}

The large scattering lengths in the two-nucleon system
require a modification of the power counting rules that follow naturally from
the single nucleon sector in order to describe multi-nucleon processes.
For processes in which all external momenta are much below the mass of the pion
($|{\bf k}|\ll m_\pi$) it is desirable to use 
an EFT without pions, which we denote by
$\nopi$.
The external momenta divided by the pion mass is
the small expansion parameter of this theory, denoted by $Q$.
As there are no dynamical pions, all quark mass effects
are reproduced  by local operators involving two or more nucleons
and an arbitrary  number of spatial derivatives.
$\nopi$ has the same power counting as 
effective range theory where the relative momentum of the nucleons
in the two-body $NN$ system is treated as the small expansion parameter.
However,  $\nopi$ includes
operators with insertions of arbitrary numbers of external currents,
that are not constrained by $NN$ scattering phase shifts alone.

Let us recall the low-momentum
behavior of the phase shift
in a system with a large scattering length.
We know from elementary scattering theory that $|{\bf k}|\cot\delta_0$ is an
analytic function of external kinetic energy  with a radius of convergence
bounded by the threshold for pion exchange in the t-channel, and thus
\begin{eqnarray}
  |{\bf k}|\cot\delta_0 & = & -{1\over a}\ +\ {1\over 2} r_0  |{\bf k}|^2
  \ +\  r_1 |{\bf k}|^4\ +\ ...
\ \ \ ,
\label{eq:pcot0}
\end{eqnarray}
where we have chosen to write the expansion in terms of the center of mass
momentum $|{\bf k}|$.
One is free to expand $|{\bf k}|\cot\delta_0$
about any point within the
region of convergence, and  eq.~(\ref{eq:pcot0}) corresponds to expanding
about $|{\bf k}|=0$.
The size of the coefficients $r_0, r_1,...$ is set by the
range of the interaction between the nucleons, of order the
pion mass, and the $r_i$ are taken to scale as $Q^0$
in the power counting.
In contrast, the scattering length $a$ is not constrained by the range
of the underlying interaction.
The scattering lengths for $NN$ scattering 
in both the $\si$ and $\siii$ channel
are very much larger than the inverse pion mass, and
are taken to scale as $a\sim Q^{-1}$
in the power counting.
In dealing with the deuteron bound state, it is convenient to make the
expansion of $|{\bf k}|\cot\delta_0$ about the location of the deuteron
pole~\cite{ERtheory}, and not about $|{\bf k}|=0$.
Expanding around the deuteron pole, $|{\bf k}|^2=-\gamma_t^2$ gives
\begin{eqnarray}
  |{\bf k}|\cot\delta_0 & = & -\gamma_t \ +\
  {1\over 2}\rho_d (|{\bf k}|^2+\gamma_t^2)\ +\
  w_2\ (|{\bf k}|^2+\gamma_t^2)^2\ +\ ...
  \nonumber\\
  & = & {\cal O}(Q) \ \ \ +\ \ \  {\cal O}(Q^2) \ \ \ \  \
  +\ \ \ \ \ \ \  {\cal O}(Q^4)\ \ \ \ \ +\ ...
\ \ \ ,
\label{eq:kcot}
\end{eqnarray}
with $\gamma_t^{-1} = 4.318946\ {\rm fm}$,
$\rho_d = 1.764\ {\rm fm}$, and $w_2=0.389\ {\rm fm^3}$\cite{Nij}.
The total energy of each nucleon in the center of mass is
$\overline{E}_N=\sqrt{|{\bf k}|^2+M_N^2}$, and the deuteron pole is located at
$|{\bf k}| = i\gamma_t$,
the solution to
\begin{eqnarray}
  -B & = & 2\sqrt{M_N^2-\gamma_t^2}  - 2 M_N
\ \ \ \ ,
\label{eq:pole}
\end{eqnarray}
where $B=2.224575~{\rm MeV}$ is the deuteron binding energy.
Subsequently, we will find it convenient to work with the parameter
$\gamma=\sqrt{M_N B}$, which is related to $\gamma_t$, by
a $Q$ expansion of eq~(\ref{eq:pole}),
\begin{eqnarray}
  \gamma & = & \gamma_t\  +\  {\gamma_t^3\over 8 M_N^2}\ +\ ...
\ \ \  .
\label{eq:gamgamt}
\end{eqnarray}

\begin{figure}[t]
\centerline{{\epsfxsize=4.5in \epsfbox{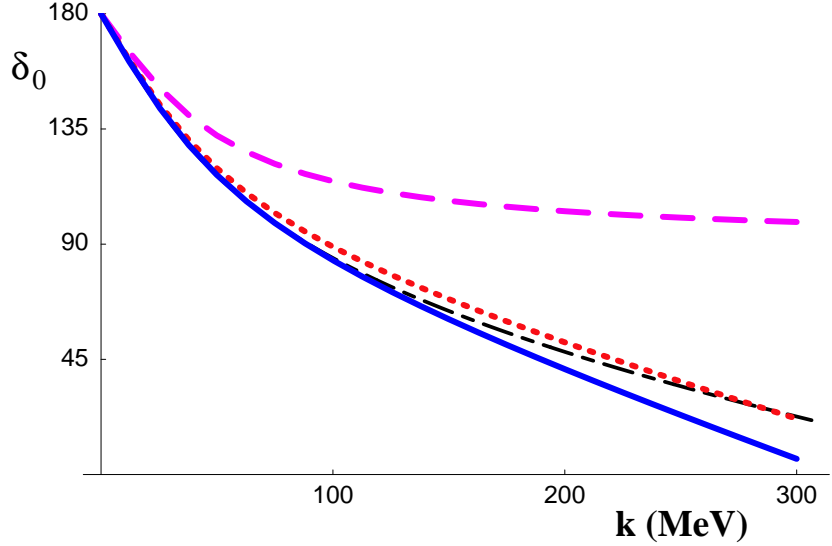}} }
\noindent
\caption{\it The phase shift $\delta_0$ as a function of the 
center of mass momentum $|{\bf k}|$. 
The dashed curve corresponds to $\delta_0^{(0)}$, 
the dotted curve corresponds to $\delta_0^{(0)}+\delta_0^{(1)}$,
the solid curve corresponds to $\delta_0^{(0)}+\delta_0^{(1)}+\delta_0^{(2)}$,
and the dot-dashed curve is the Nijmegen partial wave
analysis\protect\cite{nijmegen}.
}
\label{fig:Sphase}
\vskip .2in
\end{figure}

The S-matrix describing scattering in the coupled channel $J=1$
system is written as
\begin{eqnarray}
  S & = & \left(
    \matrix{ e^{i2\delta_0}\cos 2\overline{\varepsilon}_1
      & i e^{i(\delta_0+\delta_2)}\sin2\overline{\varepsilon}_1
      \cr
       i e^{i(\delta_0+\delta_2)} \sin2\overline{\varepsilon}_1
      &
       e^{i2\delta_2}\cos 2\overline{\varepsilon}_1
}\right)
\ \ \  ,
\label{eq:Smat}
\end{eqnarray}
where we use the ``barred'' parameterization of \cite{stapp}, also used in
\cite{Nij}.
It will follow naturally from the $\nopi$ that $\overline{\varepsilon}_1$ 
is supressed by
$Q^2$ compared with $\delta_0^{(0)}$, 
and therefore, up to N$^4$LO, we can isolate the
S-wave from the D-wave in the deuteron channel, leaving
\begin{eqnarray}
  S_{00} & = &  e^{i2\delta_0}\ =\ 1\ +\  {2 i\over \cot\delta_0 - i}
\ \ \ \ .
\label{eq:Smat0}
\end{eqnarray}
The phase shift $\delta_0$ has an expansion in powers of $Q$,
$\delta_0=\delta_0^{(0)}+\delta_0^{(1)}+\delta_0^{(2)}+...$, where the
superscript denotes the order in the $Q$ expansion.
By forming
the logarithm of both sides of eq.~(\ref{eq:Smat0}) and
expanding in powers of $Q$, it is straightforward to obtain
\begin{eqnarray}
  \delta_0^{(0)} (|{\bf k}|) & = &
  \pi - \tan^{-1}\left({ |{\bf k}|\over \gamma}\right)
  \nonumber\\
  \delta_0^{(1)}(|{\bf k}|) & = & -{\rho_d\over 2}  |{\bf k}|
  \nonumber\\
  \delta_0^{(2)}(|{\bf k}|) & = &
  -\left[ {\rho_d^2\gamma\over 4}  + {\gamma^3\over 8 M_N^2
    (\gamma^2+ |{\bf k}|^2)}\right] |{\bf k}|
  \ \ \ ,
  \label{eq:phaseexp}
\end{eqnarray}
which are shown in fig.~(\ref{fig:Sphase}).
%
Up to NNLO, the shape parameter term,
$w_2$ does not contribute to the S-wave phase shift and
therefore, in addition to
being numerically small (about a factor of 5 smaller than $\rho_d$), $w_2$
enters only at high orders.
Formally, the first deviations from linear
$|{\bf k}|$ dependence at large momenta arise from relativistic corrections
(the second term in $\delta_0^{(2)}$ in eq.~(\ref{eq:phaseexp}).
The $Q$ expansion is clearly demonstrated in eq.~(\ref{eq:phaseexp}),
by simply counting powers of $|{\bf k}|$ and $\gamma$,
which both scale like $Q$.
We expect that this perturbative expansion of the phase shifts converges
up to momenta of order $\sim m_\pi/2\sim 70~{\rm MeV}$,
at which point one encounters the t-channel cut from potential pion exchange.
In the theory where pions are dynamical, the range
of convergence of the theory is much beyond the scale set by the pion mass.
An analogous expression for the perturbatively expanded S-wave
phase shift  that is valid up to momenta of order
$\sim 300~{\rm MeV}$ at NLO
can be found in \cite{KSW}.

This simple perturbative expansion of the phase shift and
scattering amplitude
is reproduced order by order in the $\nopi$ expansion.
Relativistic corrections are encountered at NNLO and therefore a
discussion of the nucleon two-point function is appropriate.
The total energy of a nucleon $E_N$ and its kinetic energy $T_N$
are simply related by $E_N=M_N+T_N$.
A nucleon propagator can be written in terms of $T_N$
and the nucleon three momentum ${\bf p}$,
\begin{eqnarray}
  { i( \pslash+M_N) \over  p^2-M_N^2} & = &
  {2 i M_N\over (M_N+T_N)^2 - |{\bf p}|^2
    - M_N^2 + i\epsilon}
\nonumber\\
& = & 
  {2 i M_N\over 2 M_N T_N + T_N^2 - |{\bf p}|^2 + i\epsilon}
\ \ \ ,
\label{eq:prop}
\end{eqnarray}
and is reproduced by a Lagrange density involving the nucleon field $N$
with the
time-dependent phase corresponding to the nucleon mass removed,
\begin{eqnarray}
  {\cal L} & = & N^{\dagger}\left[ i\partial_0
    + {\nabla^2\over 2 M_N}\ -\
    {\partial_0^2\over 2M_N}\right] N\ +\ ...
\ \ .
\label{eq:twopt}
\end{eqnarray}
In the $\nopi$ we treat the third term in eq.~(\ref{eq:twopt}) as a
perturbation and it is not resummed into the nucleon propagator.
In order to recover the usual nonrelativistic expansion of the two-point
function, a field redefinition is employed followed by use of the equations of
motion~\cite{LukeSavage}, e.g.
\begin{eqnarray}
  N^\prime & = & \left( 1 - {\nabla^2\over 4 M_N^2} + ... \right) N
\ \ \ ,
\label{eq:fieldrd}
\end{eqnarray}
gives
\begin{eqnarray}
  {\cal L} & = & N^{\prime\dagger}\left[  i\partial_0 + {\nabla^2\over 2 M_N} +
     {\nabla^4\over 8 M_N^3}\ +\ ...\right] N^\prime \ +\ ...
\ \ \ .
\label{otherrel}
\end{eqnarray}
It is somewhat inconvenient to use this second form of the Lagrangean
due to the fact that the field redefinition in eq.~(\ref{eq:fieldrd})
must be performed on all terms in the Lagrange density describing the $NN$
systems including the multi-nucleon interactions.

The inclusion of electromagnetic interactions requires that
$\nabla$ operators be
replaced by covariant derivatives
${\bf D}=\nabla - i e {\bf A}$, and
time-derivatives $\partial_0$ be replaced by
$D_0=\partial_0+i e A_0$.
The Lagrange density describing the interaction between two nucleons scattering
in the $\siii$ channel written in terms of one-body ${\cal L}_1$ and
two-body  ${\cal L}_2$ interactions, and up to NNLO, is
 \begin{eqnarray}
{\cal L}_1 & = &  N^{\dagger}\left[ iD_0\ +\  {{\bf D} ^2\over 2 M_N}\ -\
    {D_0^2\over 2M_N}\right] N
\nonumber\\
{\cal L}_2 & = & - \Czero \left(N^T P_i N\right)^\dagger\left(N^T P_i N\right)
\nonumber\\
 & + &  \Ctwo  {1\over 8}
\left[(N^T P_i N)^\dagger
\left(N^T \left[ P_i \overrightarrow {\bf D}^2 +\overleftarrow {\bf D}^2 P_i
    - 2 \overleftarrow {\bf D} P_i \overrightarrow {\bf D} \right] N\right)
 +  h.c.\right]
\nonumber\\
 & - &  {1\over 16}\ \Cfourtwo\ 
 \left( N^T \left[ P_i \overrightarrow {\bf D}^2 +\overleftarrow {\bf D}^2 P_i
    - 2 \overleftarrow {\bf D} P_i \overrightarrow {\bf D} \right] 
    N\right)^\dagger
 \left( N^T \left[ P_i \overrightarrow {\bf D}^2 +\overleftarrow {\bf D}^2 P_i
    - 2 \overleftarrow {\bf D} P_i \overrightarrow {\bf D} \right] N\right)
\label{eq:lagtwo}
\end{eqnarray}
where $P_i$ is the spin-isospin projector for the $\siii$ channel
\begin{eqnarray}
P_i \equiv {1\over \sqrt{8}} \sigma_2\sigma_i\tau_2
\ \ \ , 
\qquad \Tr P_i^\dagger P_j ={1\over 2} \delta_{ij}
\ \ \ .
\end{eqnarray}
The subscript on the coefficient denotes the number of derivatives in the 
operator.
There is another operator involving four derivatives that we have not shown in 
eq.~(\ref{eq:lagtwo}),
 \begin{eqnarray}
{\cal L} & = & -{1\over 32}\ \Cfourone \left[ \left( N^T
(\overleftarrow {\bf D}- \overrightarrow {\bf D})^4 P_i N\right)^\dagger
\left( N^T P_i N\right)\ +\ {\rm h.c.} \right]
\ \ \ ,
\label{eq:lagsubfour}
\end{eqnarray}
(we have not shown the correct ordering of the $P_i$ and
${\bf D}$ operators).
Renormalization group scaling\cite{KSW} of the operators
in eq.~(\ref{eq:lagtwo}) and eq.~(\ref{eq:lagsubfour})
indicates that  while the contribution from 
$\Cfourtwo$ is NNLO, the contribution from $\Cfourone$ is N$^3$LO.
The time-ordered product of two $\Ctwo$ operators does not induce
the momentum structure of the $\Cfourone$ operator.

The effective range expansion provides a complete description of
scattering in the low energy region.   It is straightforward to show that
the relation between the expansion of $\cot\delta_0$ and the coefficients in
eq.~(\ref{eq:lagtwo}) is,
\begin{eqnarray}
-|{\bf k}|\cot\delta_0 & = & {2\pi \overline{E}\over M_N^2 }
{1\over \sum \Cn |{\bf k}|^{2n}}
\ +\ \mu
\ \ \ ,
\label{eq:kcotrel}
\end{eqnarray}
where $\mu$ is the dimensional regularization renormalization scale, 
explicitly introduced by the PDS subtraction procedure\cite{KSW}.
\begin{eqnarray}
\overline{E}& = & 2 M_N + \overline{T} 
= 2 M_N + { |{\bf k}|^2\over M_N } - { |{\bf k}|^4\over 4 M_N^3 }\ +\ ...
\ \ \ ,
\end{eqnarray}
is the center of mass total energy of the two nucleon system, 
where each nucleon has kinetic energy ${\overline{T}\over 2}$, and 
momentum of magnitude $|{\bf k}|$.
Lorentz invariance ensures that for two nucleons moving with 
total momentum ${\bf P}$ and kinetic energy $T$, these quantities are 
related to the center of mass quantities by
\begin{eqnarray}
T + {T^2\over 4 M_N} - {|{\bf P}|^2\over 4 M_N}  & = & 
\overline{T} + {\overline{T}^2\over 4 M_N}
\ \ \ .
\end{eqnarray}

Notice that in eq.~(\ref{eq:kcot}) momentum independent terms appear at each
order in the $Q$ expansion, as the momentum expansion
is about the deuteron pole and not about ${\bf k}=0$ (further, one sees
contributions from all powers of $|{\bf k}|^2$, up to a maximum order dictated
by $Q$).
Consequently, for the effective field theory to reproduce the effective range
expansion, the
coefficients appearing in  eq.~(\ref{eq:lagtwo}) will have an analogous
expansion in powers of $Q$,
e.g.
\begin{eqnarray}
    \Czero & = & \Czerominus + \Czerozero + \Czeroone\ +\ ...\nonumber\\
   \Ctwo & = & \Ctwozero + \Ctwoone\ +\ ...\nonumber\\
  \Cfourtwo & = & \Cfourtwoone\ +\ ...
\ \ \ .
\end{eqnarray}
The second subscript on each coefficient
denotes the powers of $Q$ in the coefficient itself.
Relating terms order by order in the ${\bf k}$ expansion of eq.~(\ref{eq:kcot})
and eq.~(\ref{eq:kcotrel}) we find that
\begin{eqnarray}
  \Czerominus & = & -{4\pi\over M_N}{1\over  (\mu-\gamma)}
  \ =\ -5.586\ {\rm fm^2}
  \nonumber\\
  \Czerozero & = & {2\pi\over M_N}{ \rho_d\gamma^2\over (\mu-\gamma)^2}
  \ =\ 0.559\ {\rm fm^2}
  \nonumber\\
  \Czeroone & = & -{\pi\over M_N} {\rho_d^2\gamma^4\over (\mu-\gamma)^3}
   \ + \ {\pi\over 2 M_N} {\gamma^3\over  M_N^2 (\mu-\gamma)^2}
  \ =\ -0.055\ {\rm fm^2}
  \nonumber\\
  \Ctwozero & = & {2\pi\over M_N}{ \rho_d\over (\mu-\gamma)^2}
  \ =\ 10.420\ {\rm fm^4}
  \nonumber\\
  \Ctwoone & = & -{2\pi\over M_N}{ \rho_d^2 \gamma^2\over (\mu-\gamma)^3}
  \ -\ {2\pi\over M_N} {1\over M_N^2 (\mu-\gamma)}
  \ =\ -2.210\ {\rm fm^4}
   \nonumber\\
  \Cfourtwoone & = & -{\pi\over M_N} {\rho_d^2\over (\mu-\gamma)^3}
  \ =\ -19.440\ {\rm fm^6}
  \ \ \ \ ,
\label{eq:Cs}
\end{eqnarray}
where the $1/M_N^2$ terms  that do not depend upon $\rho_d$
are relativistic corrections.
We have chosen to renormalize the theory at $\mu=m_\pi$.
Strictly speaking we should choose a much lower scale $\mu\sim Q$,
but the RG invariance allows us to renormalize at any scale once we have
established the power counting.
The hierarchy between coefficients of the same operator but of different orders
in the $Q$ expansion is clear. 
It is straightforward to show that these coefficients reproduce the
perturbatively expanded phase shift in eq.~(\ref{eq:phaseexp}).

The amplitude that one computes from Feynman diagrams is simply related to the
S-matrix.
Explicit computation of Feynman diagrams, in an arbitrary frame, gives
\begin{eqnarray}
  A_{NR} & = & {2\pi \overline{E} \over M_N^2}
  \ {1\over |{\bf k}|\cot\delta - i |{\bf k}|}
\ \ \ \ ,
\end{eqnarray}
which is related to the S-matrix by
\begin{eqnarray}
  S & = & 1\ +\ { 2 i\over\cot\delta-i}\ =\
  1\ +\ i {|{\bf k}| M_N^2\over \pi \overline{E} } A_{NR}
\ \ \ .
\end{eqnarray}
In the process of computing the S-matrix from the Feynman diagrams,
we have used the following S-wave states (consistent with the
sign convention of \cite{stapp})
\begin{eqnarray}
  | 1, a; {\bf P},|{\bf k}|\rangle_{L=0} & = & {1\over \sqrt{4\pi}}  
  {|{\bf k}|\over (2\pi)^3}
  \int\ d\Omega_{\bf k}
  \ \left[ N^T_{{{\bf P}\over 2} - {\bf k}}\ 
    P^a\  N_{{{\bf P}\over 2} +  {\bf k}}\right]^\dagger
\ |0\rangle
\ \ \ ,
\label{eq:swave}
\end{eqnarray}
for the $J=1, J_Z=a$ and $L=0$ NN state.
${\bf P}$ is the momentum of the center of mass of the $NN$ system.
Similarly, for the D-wave states,
\begin{eqnarray}
  | 1, a; {\bf P}, |{\bf k}|\rangle_{L=2} & = & 
  -  {3\over \sqrt{8\pi}\  |{\bf k}|}
  {1\over (2\pi)^3}
\int\ d\Omega_{\bf k}\
  \left[ {\bf k}^x {\bf k}^a - {1\over 3}\delta^{ax}  |{\bf k}|^2 \right]
  \left[ N^T_{{{\bf P}\over 2} - {\bf k}}\   P^x\  N_{{{\bf P}\over 2} + {\bf k}}
  \right]^\dagger
\ |0\rangle
\ \ \ ,
\label{eq:dwave}
\end{eqnarray}
where the $-ve$ sign gives the correct phase for the mixing parameter
$\overline{\varepsilon}_1$.
These states are normalized such that
\begin{eqnarray}
_{L^\prime}\langle 1, b; {\bf P}^\prime, |{\bf k}^\prime|
| 1, a;  {\bf P},|{\bf k}|\rangle_{L}
& = & \delta^3({\bf P}-{\bf P}^\prime)\delta(|{\bf k}|-|{\bf k}^\prime|) \delta^{ab}  
\delta^{LL^\prime}
\ \ \ .
\label{eq:norm}
\end{eqnarray}


Scattering between the S-wave and D-wave is induced by local operators first
arising at $Q^1$ in the power counting.
At order $Q^1$ and $Q^2$, the lagrange density describing such interaction is 
\begin{eqnarray}
  {\cal L}^{(sd)}_2 & = &
  {1\over 4} \CSDzero \left( N^T P^i N\right)^\dagger
  \left( N^T {\cal O}^{(sd) , xyj} N\right)
{\cal T}^{ijxy}\ +\ {\rm h.c.}
\nonumber\\
& - & {1\over 16}\CSDtwotwo
\left( N^T \left[ P_i \overrightarrow {\bf D}^2 +\overleftarrow {\bf D}^2 P_i
- 2 \overleftarrow {\bf D} P_i \overrightarrow {\bf D} \right]  N\right)^\dagger
  \left( N^T {\cal O}^{(sd) , xyj} N\right)
{\cal T}^{ijxy}\ +\ {\rm h.c.}
\ \ \ ,
\label{eq:sdlag}
\end{eqnarray}
where
\begin{eqnarray}
{\cal O}^{(sd) , xyj} & = &
      \overleftarrow {\bf D}^x  \overleftarrow {\bf D}^y P^j
+ P^j \overrightarrow {\bf D}^x  \overrightarrow {\bf D}^y
- \overleftarrow {\bf D}^x P^j\overrightarrow {\bf D}^y
-\overleftarrow {\bf D}^y P^j\overrightarrow {\bf D}^x
\nonumber\\
{\cal T}^{ijxy}
& = & 
\left( \delta^{ix}\delta^{jy} - {1\over n-1} \delta^{ij}\delta^{xy}\right)
\ \ \ ,
\label{eq:sdop}
\end{eqnarray}
and where $n$ is the number of spacetime dimensions.
When electromagnetic interactions are ignored the operator in
eq.~(\ref{eq:sdop}) collapses to
\begin{eqnarray}
{\cal O}^{(sd) , xyj}&\rightarrow &
P^j  \left( \overleftarrow {\bf \nabla} - \overrightarrow {\bf \nabla}\right)^x
 \left( \overleftarrow {\bf \nabla} - \overrightarrow {\bf \nabla}\right)^y
\ \ \ ,
\end{eqnarray}
which is explicitly Galilean invariant.
As the  S-D mixing parameter is being computed up to NLO relativistic
corrections do not enter.
There is another operator with four derivatives that contributes
to S-D mixing, 
\begin{eqnarray}
  \tilde {\cal L}^{(sd)}_2 & = &
- {1\over 16}\CSDtwoone  \left( N^T P^i N\right)^\dagger
  \left( N^T \left[
( \overleftarrow {\bf D} )^2 {\cal O}^{(sd) , xyj}
+  {\cal O}^{(sd) , xyj} ( \overrightarrow {\bf D} )^2
- 2 \overleftarrow {\bf D}^l {\cal O}^{(sd) , xyj} \overrightarrow {\bf D}^l
\right]
N\right)
{\cal T}^{ijxy}
\ \ \ ,
\label{eq:sdlagsub}
\end{eqnarray}
(along with its hermitian conjugate)
but it is of order $Q^3$  and does not contribute at NLO.

\begin{figure}[t]
\centerline{{\epsfxsize=4.5in \epsfbox{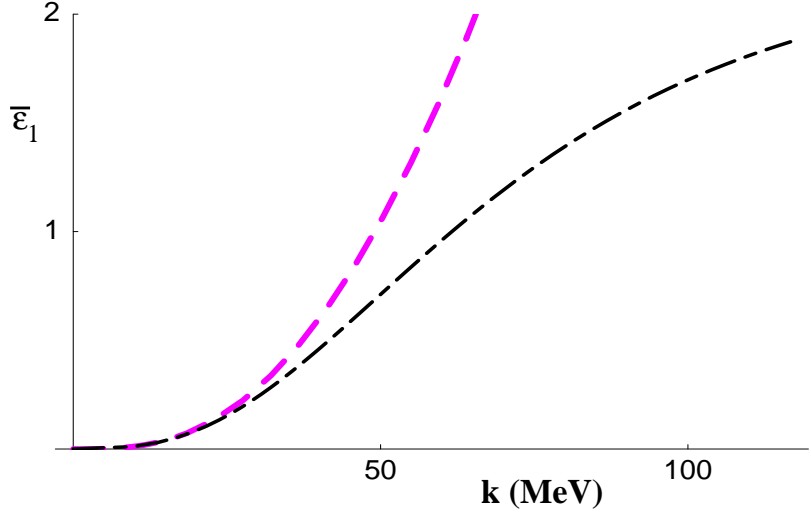}} }
\noindent
\caption{\it The S-D mixing parameter $\overline{\varepsilon}_1$,
in degrees,
as a function of the 
center of mass momentum $|{\bf k}|$. 
The dashed curve corresponds to $\overline{\varepsilon}_1^{(2)}$, 
and the dot-dashed curve is the Nijmegen partial wave analysis.
}
\label{fig:Epphase}
\vskip .2in
\end{figure}

The coefficients that appear in 
eq.~(\ref{eq:sdop})
themselves have an expansion in powers of $Q$,
e.g. $\CSDzero = \CSDzeroone + \CSDzerotwo+...$.
Performing a $Q$ expansion on the mixing parameter
$\overline{\varepsilon}_1 =
\overline{\varepsilon}^{(2)}_1
+\overline{\varepsilon}^{(3)}_1+...$
it is straightforward to demonstrate that
\begin{eqnarray}
\overline{\varepsilon}_1^{(2)} (|{\bf k}|) & = &
 {\sqrt{2}\over 3} \left({  \CSDzeroone \over \Czerominus  }\right)
{ |{\bf k}|^3\over\sqrt{\gamma^2+|{\bf k}|^2}}
\nonumber\\
& = &
-{M_N\over 4\pi} (\mu-\gamma)  {\sqrt{2}\over 3}\ 
\CSDzeroone  { |{\bf k}|^3\over\sqrt{\gamma^2+|{\bf k}|^2}}
\ \ \  .
\label{eq:eonelead}
\end{eqnarray}
Renormalization group invariance of this leading order contribution to
$\overline{\varepsilon}_1$ indicates that
$\CSDzeroone\propto (\mu-\gamma)^{-1}$.
Renormalizing at the scale $\mu=m_\pi$, and comparing with the Nijmegen 
phase shift
analysis\cite{nijmegen}, we find $\CSDzeroone=-4.57\ {\rm fm}^4$.
A comparison between the Nijmegen phase shift analysis and our fit can be
seen in fig.~(\ref{fig:Epphase}).

%

At the next order, $Q^3$, the contribution to $\overline{\varepsilon}_1$ is
\begin{eqnarray}
 \overline{\varepsilon}_1^{(3)}(|{\bf k}|) & = &  -{\sqrt{2}\over 3} {M_N \over 4\pi} 
 \left[
\left( {\mu\gamma\rho_d \over 2}  \ \CSDzeroone
  \ +\  (\mu-\gamma)\ \CSDzerotwo \right)
{ |{\bf k}|^3\over\sqrt{\gamma^2+|{\bf k}|^2}}
\right. \nonumber\\
& & \left.   
\qquad   \ +\
\left( {\rho_d\over 2}\   \CSDzeroone + (\mu-\gamma)\  \CSDtwotwotwo\ \right)
{ |{\bf k}|^5\over\sqrt{\gamma^2+|{\bf k}|^2}}
   \right]
\ \ \ ,
\label{eq:eonesub}
\end{eqnarray}
where we have expanded $\CSDtwotwo=\CSDtwotwotwo+...$.
The first term in eq.~(\ref{eq:eonesub}) has the same momentum dependence as
$\overline{\varepsilon}_1^{(2)}$, the
leading contribution in eq.~(\ref{eq:eonelead}).
Requiring that the momentum structure of $\overline{\varepsilon}_1^{(2)}$ does not
appear at higher orders in the expansion constrains $\CSDzerotwo$ to be 
\begin{eqnarray}
 \CSDzerotwo & = &  -{\mu\gamma\rho_d \over 2 (\mu-\gamma)}  \ \CSDzeroone
\ \ \ .
\label{eq:csdtwocon}
\end{eqnarray}
Further, as the second term in eq.~(\ref{eq:eonesub})
is an observable, we obtain the RG equation
\begin{eqnarray}
\mu {d\over d\mu} 
\left( {\rho_d\over 2}\   \CSDzeroone + (\mu-\gamma)\  \CSDtwotwotwo\  \right)
& = & 0
\ \ \ .
\label{eq:eoneRG}
\end{eqnarray}
By considering $\mu$ to be of order the  short-distance matching scale it 
is clear that the combination 
${1\over 2}\rho_d \   \CSDzeroone + (\mu-\gamma)\  \CSDtwotwotwo\ $ 
appearing in eq.~(\ref{eq:eonesub}) and eq.~(\ref{eq:eoneRG})
is of order $Q^0$ and not $Q^{-1}$, as naively guessed from the subscripts
of $ \CSDzeroone$ and $\CSDtwotwotwo$.
Therefore 
\begin{eqnarray}
 \overline{\varepsilon}_1^{(3)}(|{\bf k}|) & = & 0
\ \ \ ,
\label{eq:ep1zero}
\end{eqnarray}
and the first corrections to $\overline{\varepsilon}_1^{(2)}$
arise at NNLO\footnote{This was pointed out to us by S. Fleming, T. Mehen and 
I. Stewart, and can be found in their recent preprint\cite{FMSep}.},
i.e.  $\overline{\varepsilon}_1^{(4)}$ .

Local operators that contribute to D-wave to D-wave scattering contain
four or more powers of the external momenta.
A RG analysis shows that the coefficient of the leading interaction scales like
$\mu^{-1}$, through the time-ordered product of two $\CSDzero$ operators.
Therefore, contributions to the scattering amplitude start at order $Q^3$
(N$^4$LO, i.e. $\delta_2^{(4)}$ is the first non-zero contribution to the
$\delta_2$ phase shift), which is higher order than we are working.


\section{Properties of the Deuteron}

In the previous section nucleon-nucleon scattering was used to determine the
coefficients that appear in $\nopi$  up to NNLO for S-wave
interactions and up to NLO for mixing between the S-wave and
D-wave.
This provides us with enough information to determine several
properties of the deuteron up to NNLO in $\nopi$.

To begin with it is informative to recover basic kinematic properties of the
deuteron from the effective field theory formalism.
The inclusion of the deuteron into EFT
was discussed in detail in \cite{KSW2}, but relativistic corrections were not
included.
In order to compute the matrix element of the electromagnetic current between
deuteron states, one starts by defining the 3-point  function
for two nucleons with kinetic energy $T$ interacting
with the electromagnetic current, transferring four-momentum $q$ and
producing two nucleons with kinetic energy $T^\prime$,
\begin{eqnarray}
  G^\mu_{ij} (T,T^\prime, q) & = &
  \int\ d^4x\ d^4y\
  e^{+i (T^\prime x_0 - {\bf p}^\prime\cdot{\bf x})} 
  e^{-i (T y_0 - {\bf p} \cdot{\bf y})} 
\langle 0| T[ {\cal D}_i (x) J^\mu_{\rm em} (0)  {\cal D}_j (y)^\dagger ]
 0\rangle
\ \ \ .
\label{eq:Tord}
\end{eqnarray}
where $ {\cal D}_i (x) = N^T P_i N (x)$.
The S-matrix element for this interaction is found by LSZ reduction of this
Green-function
\begin{eqnarray}
  \langle  {\bf p}^\prime, j| J^\mu_{\rm em} | {\bf p},  i \rangle
  & = &
\left(\sqrt{Z}\right)^2
\left[ G^{-1} (T)  G^{-1} (T^\prime)
  G^\mu_{ij} (T,T^\prime, q)
  \right]_{T\rightarrow T_{\rm pole} ,
    T^\prime \rightarrow T^\prime_{\rm pole}}
\ \ \ ,
\label{eq:Sdef}
\end{eqnarray}
where $T_{\rm pole}$ is the location of the deuteron pole for two nucleons
moving with three-momentum ${\bf p}$,
\begin{eqnarray}
 &  &  T_{\rm pole}\ +\ { T_{\rm pole}^2\over 4 M_N} - {|{\bf p}|^2\over 4 M_N} \
 =\ -{\gamma^2\over M_N} + {\gamma^4\over 4 M_N^3}
\nonumber\\
 &  &  T_{\rm pole}\ =\ -{\gamma^2\over M_N} \ + { |{\bf p}|^2\over 4 M_N}\ +\
 {\gamma^2  |{\bf p}|^2\over 8 M_N^3}\ -\  { |{\bf p}|^4\over 64 M_N^3}
\ +\ ...\ \ .
\label{eq:poleloc}
\end{eqnarray}
The expression for $T_{\rm pole}$ has been expanded in powers of $M_N$
because it is the nucleon mass that will naturally arise from
Feynman diagrams computed with the effective field theory.
An analogous expression holds for  $T_{\rm pole}^\prime $.
The ellipses in eq.~(\ref{eq:poleloc}) denote terms of order $Q^6$ or higher.
The two nucleon two-point function, defined in $\cite{KSW2}$ can be generalized
to include relativistic effects,
\begin{eqnarray}
  G(T) & = & \int\ d^4x \ e^{i(T x_0-{\bf p}\cdot {\bf x})}
\langle 0| T[ {\cal D}_i (x) {\cal D}_j^\dagger (0)]
 0\rangle
 \ =\
 \delta_{ij} {i {\cal Z} (T)\over T-T_{\rm pole} + i\epsilon}
\ \ \ .
\label{eq:twoptgreen}
\end{eqnarray}
The wavefunction renormalization factors are straightforward to compute from a
single nucleon-nucleon bubble, which is explicitly lorentz invariant (up to the
order one computes).
An important modification arising from the inclusion of relativistic
effects is that the two-point function
is an explicit function of
$ T + {T^2\over 4 M_N } - { |{\bf p}|^2\over 4 M_N}$.
This leads to wavefunction renormalization factors of 
\begin{eqnarray}
  Z & = & {\cal Z} (T_{\rm pole})\ =\ 
  \left[ {  1 + {T_{\rm pole}\over 2 M_N} \over d\Sigma (T)/  dT}
    \right]_{T\rightarrow T_{\rm pole} }
\ \ \ ,
\label{eq:zpsi}
\end{eqnarray}
where $\Sigma (T)$ is the irreducible bubble presented in \cite{KSW2}.
We find that
\begin{eqnarray}
  \Zpi & = & -{8\pi\gamma\over M_N^2} 
\left[\  1 \ +\  \gamma \rho_d\ +\ \gamma^2 \rho_d^2 \ + \
  {\gamma^2 (7\mu-5\gamma)\over 8 M_N^2 (\mu-\gamma)} \ +\ ...\ 
  \right]
\ \ \ ,
\label{eq:zpsiexp}
\end{eqnarray}
which is lorentz invariant, but depends on the
renormalization scale $\mu$.

Evaluating the matrix element of the Hamiltonian between deuteron states
\begin{eqnarray}
  \langle  {\bf p}^\prime , j| H | {\bf p},  i \rangle
  & = & 
  \langle  {\bf p}^\prime, j| \int d^3 x \left[ N^\dagger \left( M_N\ +\
      {\nabla^2\over 2M_N} \right) N
    \ +\ \Czero \left(N^T P_i N\right)^\dagger \left(N^T P_i N\right)
  \ +\ ...\right]
  | {\bf p},  i \rangle
\nonumber\\
& = &  \left( 2 M_N\ +\ {|{\bf p}|^2\over 4 M_N} 
\ -\ {\gamma^2\over M_N}\ +\ ...\right) 
(2\pi)^3\delta^3({\bf p}-{\bf p}^\prime)\delta_{ij}
\nonumber\\
& = &  \left( M_d \ +\ {|{\bf p}|^2\over 2 M_d}\ +\ ...\right) 
(2\pi)^3\delta^3({\bf p}-{\bf p}^\prime)\delta_{ij}
\ \ \ .
\label{eq:EP}
\end{eqnarray}
The interactions and relativistic contributions combine together in
such a way to reproduce the deuteron energy-momentum relation at the order to
which the calculation is being performed.
This is a generic feature, which must be present for the theory to be
consistent.  In computing any observable, one finds that the deuteron mass is
recovered order by order in perturbation theory.
The issue of gauge invariance for the deuteron bound state is also simply
addressed.
The lorentz-invariant
lagrange density that describes the leading low-energy interactions of the deuteron
has the form
\begin{eqnarray}
  {\cal L} & = &
  d_j^\dagger \left[
    iD_0\ +\ { {\bf D}^2\over 2 M_d}\ -\ { D_0^2\over 2 M_d}\right] d_j
  \ +\ ...
\ \ \ ,
\label{eq:deutlag}
\end{eqnarray}
and so reproducing the correct energy-momentum relation for the deuteron order
by order in the expansion forces the matrix elements to be gauge invariant
order by order in the expansion.
In fact, direct calculation including the leading relativistic corrections gives
\begin{eqnarray}
  {\cal L} & = &
  d_j^\dagger \left[
    i (\partial_0 + i e A_0) 
    \ +\ ( {\bf\nabla}- i e {\bf A})^2\left( {1\over 4 M_N} + {\gamma^2\over 8
        M_N^3}\right)
    \ -\ (\partial_0 + i e A_0)^2 \left( {1\over 4 M_N}\right)
  \right] d_j
  \ +\ ...
\ \ \ .
\label{eq:deutmatch}
\end{eqnarray}
Order by order one recovers the correct
matrix elements of $J^\mu_{\rm em}$ to reproduce the
couplings induced by eq.~(\ref{eq:deutlag}) and also the Thompson limit
for photon-deuteron elastic scattering.


\subsection{Electric Polarizability of the Deuteron}

The electric polarizability of the deuteron, $\alpha_{E0}$,
has been investigated thoroughly
with potential models\cite{FFa}-\cite{Khar}.
To a very high precision one obtains
$\alpha_{E0}=0.6328\pm 0.0017~{\rm fm}^3$
with those techniques.
This is no surprise because the electric polarizability is dominated by the
long range behavior of the deuteron wavefunction.
If the model is tuned to reproduce this component of the wavefunction the
predicted electric polarizability
should be very close to nature.
In effective range theory the polarizabilities are assumed to
be dominated by the asymptotic S-wave component of the
deuteron wave function,
\begin{eqnarray}
  \psi^{({\rm ER})}({\bf r}) & = & \sqrt{{\gamma \over 2 \pi
 (1- \gamma \rho_d) }} {e^{-\gamma r} \over r}
  \ \ \ ,
  \label{eq:ERwave}
\end{eqnarray}
which yields an electric polarizability of\cite{FFa,LucasRust}
\begin{eqnarray}
   \alpha_{E0}^{\rm ER}  & = & {\alpha M_N\over 32\gamma^4} {1\over 1 - \gamma
     \rho_d}
   \nonumber\\
    & = & {\alpha M_N\over 32\gamma^4}
    \left[ 1 + \gamma \rho_d\ + \gamma^2 \rho_d^2\ +\
      ...\right]
    \ \ \ ,
\label{eq:ERpol}
\end{eqnarray}
where $\alpha$ is the electromagnetic fine structure constant.
Numerically, $\alpha_{E0}^{\rm ER}=0.6338~{\rm fm}^3$,
which is very close to
the value obtained by the potential model calculations.

With the $\nopi$ we have computed  $\alpha_{E0}$ up to NNLO, including
relativistic corrections.   As with the phase shifts, it has a perturbative
expansion in $Q$, 
$\alpha_{E0} = \alpha_{E0}^{(-4)}+ \alpha_{E0}^{(-3)} + \alpha_{E0}^{(-2)}+...$,
and we find
\begin{eqnarray}
     \alpha_{E0}^{(-4)}+ \alpha_{E0}^{(-3)} + \alpha_{E0}^{(-2)}
     & = & 
{\alpha M_N\over 32\gamma^4}
    \left[ 1 \ \ +\ \  \gamma \rho_d\ \ +\ \  \gamma^2 \rho_d^2
      \ \ +\ \ {2\gamma^2\over 3 M_N^2}
      \right]
\nonumber\\
& = & \qquad 0.377\ +\ 0.153\ +\ 0.062\ +\ 0.0006\
\nonumber\\
& = & 0.592~{\rm fm}^3
\ \ \ .
\label{eq:eftpol}
\end{eqnarray}
Numerically, the relativistic corrections are very small, two orders of
magnitude smaller than the NNLO corrections from the four-nucleon
interactions.
The value of $\alpha_{E0}$ shown in eq.~(\ref{eq:eftpol})
is within $\sim 5\%$ of that computed with potential models
and with effective range theory.
Despite making a small contribution to the numerical value of  $\alpha_{E0}$
this calculation demonstrates that relativistic contributions can be computed
easily with the EFT.
The operators that mix S-wave and D-wave
(corresponding to the D-wave component of the deuteron in potential model
language)
make contributions to   $\alpha_{E}^{(-2)}$.
They do not contribute to the scalar polarizability,
$\alpha_{E0}^{(-2)}$, but do  contribute to the 
tensor polarizability, $\alpha_{E2}^{(-2)}$. 
Such operators will contribute to $\alpha_{E0}$ 
at higher orders in the expansion.

The four-nucleon operators in the
$\nopi$ reproduce the contributions from  effective range theory
in addition to
the relativistic corrections that arise at NNLO.
However, the situation is different at  higher orders.
$\alpha_{E0}^{(-1)}$ will receive contributions from P-wave 
interaction between nucleons
(a two-derivative operator that is not renormalized by the large S-wave
scattering length).
$\alpha_{E0}^{(0)}$ receives contributions 
from the polarizability of the nucleons themselves.
It is only $\alpha_{E0}^{(1)}$ and higher that receive contributions 
form the
four-nucleon-two-photon interaction, a counterterm for the polarizability of
the deuteron, that can only be determined from two-nucleon and two-photon
processes.
Extrapolating the relative size of the contributions seen in
eq.~(\ref{eq:eftpol})
we conclude that the counterterm contribution will be of order
$\sim 0.005$, much larger than the relativistic corrections, but still only
$\sim 1\%$.

In the EFT with pions included as a dynamical field, the electric
polarizability at NLO is\cite{CGSSpol}
\begin{eqnarray}
  \alpha_{E0} & = & {\alpha M_N\over 32\gamma^4}
  \left[
    1
    \ +\ C_2(\mu)\ {M_N \gamma (\mu-\gamma)^2\over 2\pi }
    \ +\ {g_A^2 M_N \gamma m_\pi^2 (3 m_\pi^2+16 m_\pi \gamma + 24\gamma^2)
    \over 12 \pi f^2 (m_\pi + 2\gamma)^4}
    \right]
\nonumber\\
 & \rightarrow &
{\alpha M_N\over 32\gamma^4}
  \left[
    1
    \ +\ C_2(\mu)\ {M_N \gamma (\mu-\gamma)^2\over 2\pi }
    \ +\ { g_A^2 M_N \gamma \over 4 \pi f^2} 
    \left( 1 - {8 \over 3}{\gamma\over m_\pi} + {16\over 3}
      {\gamma^2\over m_\pi^2}\ +\ ...\right) 
    \right]
\nonumber\\
 & = &
{\alpha M_N\over 32\gamma^4}
  \left[ 1 \ +\  \gamma \rho_d
  \ +\ { g_A^2 M_N \gamma \over 4 \pi f^2}{10\over 3}
      {\gamma^2\over m_\pi^2}\ +\ ...
      \right] 
\ \ \ ,
\label{eq:polER}
\end{eqnarray}
where $g_A$ is the pion-nucleon axial coupling constant, and $f$ is the pion
decay constant.
From the expansion of $\alpha_{E0}$
in powers of $\gamma/m_\pi$
that appears in eq.~(\ref{eq:polER}) it is clear that one recovers the effective
range expansion, but with an additional contribution from the pions that is
higher order in the momentum expansion.
In the $\nopi$, $\gamma/m_\pi\sim Q$ while in the theory with dynamical pions
it is $\gamma/m_\pi\sim Q^0$.
The residual contribution on the third line of  eq.~(\ref{eq:polER})
proportional to $g_A^2$ is
therefore N$^3$LO in the momentum expansion appropriate to the theory
without pions.


\subsection{Electric Charge Form-Factor of the Deuteron}

The electric charge form factor of the deuteron is a very well measured object
over a wide range of momentum transfers.
Potential models and effective range theory reproduce the data very well in the
kinematic regions where they are applicable.
(For a recent and very comprehensive review, see \cite{Wong}.)

A deuteron with four-momentum $p^\mu$ and polarization vector
$\epsilon^\mu$ is
described by the state $\ket{\bfp ,\epsilon}$, where the polarization vector
satisfies $p_\mu\epsilon^\mu=0$.  An orthonormal basis of
polarization vectors $\epsilon_i^\mu$
satisfies
\begin{eqnarray}
p_\mu\epsilon_i^\mu=0\ ,\qquad \epsilon_{i\mu}^*\epsilon^\mu_j =- \delta_{ij}\
,\qquad \sum_{i=1}^3\epsilon^{*\mu}_i\epsilon^\nu_i = {p^\mu p^\nu\over
M_d^2}-g^{\mu\nu}
\ \ \ \ ,
\label{eq:spindef}
\end{eqnarray}
where $M_d$ is the deuteron mass.
It is convenient to choose the basis polarization vectors so that in
the deuteron rest frame $\epsilon_i^\mu=\delta^{\mu}_i$.
Deuteron states with these polarizations are denoted by
$\ket{\bfp,i}$ (i.e., $\ket{\bfp,i} \equiv \ket{\bfp,\epsilon^\mu_i}$)
and satisfy the normalization condition
\begin{eqnarray}
\langle\bfpp,j\vert\bfp,i\rangle =(2\pi)^3\delta^3(\bfp-\bfpp)
\delta_{ij}
\ \ \ .
\end{eqnarray}
In terms of these states the nonrelativistic expansion of 
the matrix element of the electromagnetic current is (up to NNLO)
\begin{eqnarray}
\bra{\bfpp,j} J^0_{em}\ket{\bfp,i} &=& e \[  F_C(q^2) \delta_{ij} + {1\over 2
M_d^2}F_{\cal Q}
(q^2)\(\bfq_i\bfq_j-{1\over n-1}\bfq^2 \delta_{ij}\)\]
\left({E+E^\prime\over 2 M_d}\right)
\ \ \ ,
\nonumber\\
\bra{\bfpp,j} {\bf J}^k_{em}\ket{\bfp,i}&=& {e\over 2 M_d} \[ F_C(q^2)
\delta_{ij}(\bfp+\bfpp)^k + F_M(q^2)\(\delta_j^k\bfq_i - \delta_i^k\bfq_j\)
\right.\nonumber\\
&&\qquad\quad \left.+{1\over 2M_d^2} F_{\cal Q}(q^2) \(\bfq_i\bfq_j-{1\over
n-1}\bfq^2\delta_{ij}\)(\bfp+\bfpp)^k\]
 \ \ \ ,
\label{eq:emmatdef}
\end{eqnarray}
where
${\bf q}={\bf p}^{\prime}-{\bf p}$,
$q^2=q_0^2-|{\bf q}|^2$,
is the square of the four-momentum transfer, and $n$ is the number of
space-time dimensions.
(Note that we are using a different normalization of states
to \cite{KSW2,FriarEM}.).
The dimensionless form factors defined in eq.~(\ref{eq:emmatdef})
are normalized such that
\begin{eqnarray}
F_C(0) &=& 1\ \ \ ,
\nonumber\\
{e\over 2 M_d}F_M(0) &=& \mu_{M}\ \ \ ,
\nonumber\\
{1 \over M_d^2} F_{\cal Q}(0) &=& \mu_{\cal Q}\ ,
\label{eq:normalization}
\end{eqnarray}
where $\mu_M= 0.85741 \ {e\over 2M_N}$ is the
deuteron magnetic moment,
and $\mu_{\cal
Q}=0.2859\,{\rm fm}^2$ is the
deuteron quadrupole moment.
The charge radius of the deuteron $\sqrt{\langle r_d^2\rangle}$
is defined by
\begin{eqnarray}
F_C (q^2) & = & 1\  +\  {1\over 6} \langle r_d^2\rangle\  q^2\  +\ ... 
\ \ \ .
\label{eq:raddef}
\end{eqnarray}

In effective range theory, the short-distance part of the deuteron
wave function is only important
for establishing the charge normalization condition, $F_C(0)=1$.
The prediction of
effective range theory for the form factor $F_C(q^2)$ follows from the
Fourier transform of $ |\psi^{(ER)}({\bf r})|^2$,
\begin{eqnarray}
F^{(ER)}_C(q^2)=\left({1 \over 1- \gamma \rho_d} \right)
\left(\left(
{4 \gamma \over \sqrt{-q^2}}\right){\rm tan}^{-1}\left({\sqrt{-q^2} \over 4
  \gamma}\right) -\gamma\rho_d\right)
\ \ \ .
\label{eq:ERcff}
\end{eqnarray}
This yields a deuteron charge radius,
\begin{eqnarray}
   \langle r_d^2\rangle^{\rm ER}  & = & {1\over 8\gamma^2} {1\over 1 - \gamma
     \rho_d}
   \nonumber\\
    & = & {1\over 8\gamma^2}\left[ 1 + \gamma \rho_d\ + \gamma^2 \rho_d^2\ +\
      ...\right]
    \ \ \ ,
\label{eq:ERcr}
\end{eqnarray}
which gives a numerical value of
$\sqrt{ \langle r_d^2\rangle^{\rm ER}}=1.98~{\rm fm}$,
very close to the
quoted value of the matter radius of the deuteron\cite{buch,FMS} of
$r_m=1.967\pm 0.002~{\rm  fm}$.
Conventionally, the charge radius is obtained by combining
the nucleon charge radius in quadrature with the matter radius, and agrees
very well with the experimental value.

In the $\nopi$, the charge form factor $F_C (q^2)$ has
an expansion in powers of $Q$,
$F_C (q^2) =F_C^{(0)} (q^2)
+F_C^{(1)} (q^2)+F_C^{(2)} (q^2)+...$
where the superscript denotes the order of the contribution.
The leading contribution to charge form factor $F_C (q^2)$
is calculated to be
\begin{eqnarray}
  F_C^{(0)} (q^2) & = & {4\gamma\over\sqrt{-q^2}} \tan^{-1}\left(
    {\sqrt{-q^2}\over 4\gamma}\right)
\ \ \ ,
\label{eq:EFTcffLO}
\end{eqnarray}
which reproduces the leading term in a $\gamma\rho_d$ expansion of the
form factor computed with effective range, eq.~(\ref{eq:ERcff}).
At NLO, we find a contribution to the charge form-factor of
\begin{eqnarray}
  F_C^{(1)} (q^2) & = & - \gamma\rho_d \left[ 1 - 
{4\gamma\over\sqrt{-q^2}} \tan^{-1}\left( {\sqrt{-q^2}\over 4\gamma}\right)
\right]
\ \ \ ,
\label{eq:EFTcffNLO}
\end{eqnarray}
which reproduces the subleading term in the 
$\gamma\rho_d$ expansion of the
effective range result, eq.~(\ref{eq:ERcff}).
At NNLO, there are contributions from several types of graphs, including
relativistic corrections.  We find
\begin{eqnarray}
  F_C^{(2)} (q^2) & = & 
- \gamma^2\rho_d^2 \left[ 1 - 
{4\gamma\over\sqrt{-q^2}} \tan^{-1}\left( {\sqrt{-q^2}\over 4\gamma}\right)
\right]
\nonumber\\
& + &
{1\over M_N^2} \left[
  {10\gamma^4\over 16\gamma^2 - q^2}
\ -\ {\gamma^2\over 2}
  \ -\
{1\over 32}\left(q^2 + 4 \gamma^2\right)
  {4\gamma\over\sqrt{-q^2}} \tan^{-1}\left( {\sqrt{-q^2}\over 4\gamma}\right)
\right]
\nonumber\\
& + &
{1\over 6} \langle r_{N,0}^2\rangle \ q^2\ 
{4\gamma\over\sqrt{-q^2}} \tan^{-1}\left(
    {\sqrt{-q^2}\over 4\gamma}\right)
  \ \ \ ,
\label{eq:EFTcffNNLO}
\end{eqnarray}
where the first line of eq.~(\ref{eq:EFTcffNNLO}) reproduces the subsubleading
term in the effective range expansion, eq.~(\ref{eq:ERcff}), while the second
term is the relativistic correction.
The third term is the contribution from the
isoscalar charge radius of the nucleon,
$\sqrt{ \langle r_{N,0}^2\rangle}$,
which is measured to be
$\sqrt{ \langle r_{N,0}^2\rangle} = 0.79\pm 0.01~{\rm fm}$.
This measured value contains both relativistic corrections,
$\sqrt{ \langle r_{N, {\rm rel}}^2\rangle} = \sqrt{3\over 4 M_N^2} = 0.18~{\rm fm}$
(the Foldy term), and contributions from strong interactions.
There is no reason to separate these two contributions,
as $\langle r_{N,0}^2\rangle$ is the coefficient of
the nucleon-photon
charge radius operator in the single nucleon sector.

The charge radius of the deuteron resulting from this form factor is
\begin{eqnarray}
 \langle r_d^2\rangle^{\rm EFT} & = &\langle r_{N,0}^2\rangle
  \ +\ 
 {1\over 8\gamma^2}
 \left[ 1\ +\ \gamma\rho_d\ +\ \gamma^2\rho_d^2\right]
   \ +\ {1\over 32 M_N^2}
   \nonumber\\
& = & 0.62\ \ \ \  +\ \ \ \ 2.33\ +\ 0.95\ +\ 0.39 \ +\   0.0014
\nonumber\\
& = & 4.30~{\rm fm}^2
\ \ \ \ ,
\label{eq:EFTcr}
\end{eqnarray}
where the last term is the relativistic correction.
Taking the square root of this value gives
$\sqrt{\langle r_d^2\rangle}=2.07~{\rm fm}$,
which is within a few percent of the measured value of
$\sqrt{\langle r_d^2\rangle}=2.1303~{\rm fm}$\cite{Wong,buch,FMS,EWa}.
The magnitude of the relativistic correction we have computed, 
$\langle r^2_{d, {\rm rel} }\rangle=+0.0014~{\rm fm}^2$, 
is the same as the contribution from the spin-orbit 
interaction (relativistic effect) computed in \cite{FMS}, 
but is of opposite sign.

A comparison between eq.~(\ref{eq:EFTcr})
and 
eq.~(\ref{eq:ERcr}) reveals that charge radius computed up to NNLO in
$\nopi$ is the same as that computed with effective range theory up to
very small relativistic corrections.
However, at N$^3$LO
there is a contribution to the charge form factor and to the
charge radius from a four-nucleon-one-photon operator of the form
\begin{eqnarray}
  {\cal O}_{\rm ct} & = &
  \left( N^T P_i N\right)^\dagger \left( N^T P_i N\right) \nabla^2 A_0
\ \ \ .
\end{eqnarray}
Therefore at N$^3$LO the prediction of $\nopi$ and effective range theory will
differ.


\subsection{Magnetic Form-Factor of the Deuteron}

The magnetic form-factor of the deuteron has been computed in the effective field
theory with dynamical pions\cite{KSW2} and is found to be dominated  by the
nucleon isoscalar magnetic moment.
At NLO there are no mesonic corrections other than those that can be absorbed
into the definition of the nucleon magnetic moment.
However, there is a contribution from a four-nucleon-one-photon counterterm
with an unknown coefficient.
As the pions play no explicit role in the deuteron magnetic moment at NLO,
one can immediately write an expression for the magnetic moment and form factor
in $\nopi$ from the expressions given in \cite{KSW2}.

The Lagrange density describing the magnetic interactions of the nucleons is
\begin{eqnarray}
{\cal L}_{1,B} & = &
{e\over 2 M_N} N^\dagger
\left( \kappa_0 + \kappa_1 \tau_3 \right) {\bf \sigma} \cdot {\bf B} N
\ \ \ \ ,
\label{eq:Nucmag}
\end{eqnarray}
where
$\kappa_0 = {1\over 2} (\kappa_p + \kappa_n)$ and
$\kappa_1 = {1\over 2} (\kappa_p - \kappa_n)$
are the isoscalar and isovector nucleon magnetic moments in nuclear magnetons, 
with
\begin{eqnarray}
\kappa_p & =&  2.79285\ ,\qquad\kappa_n = - 1.91304
\ \ \ \ .
\label{eq:nucmagdef}
\end{eqnarray}
The magnetic field  is conventionally defined 
${\bf B} ={\bf \nabla} \times {\bf A}$.
At NLO there are four-nucleon-one-photon operators that appear and can
contribute to both the deuteron magnetic moment and the rate for
$np\rightarrow d\gamma$,
\begin{eqnarray}
{\cal L}_{2,B}
& = &  \left[e\  \Lone \ (N^T P_i N)^\dagger (N^T \overline{P}_3 N) B_i
\ -\ 
 e\  \Ltwo \  i\epsilon_{ijk} (N^T P_i N)^\dagger (N^T P_j N)  B_k
+ {\rm h.c.} \right]
\ \ \ \ .
\label{eq:Ldef}
\end{eqnarray}
At NLO the deuteron magnetic moment is
\begin{eqnarray}
  \mu_M & = & {e\over 2 M_N} \left(\kappa_p\ +\ \kappa_n\ +\ \Ltwo\ { 2 M_N \gamma
    (\mu-\gamma)^2\over\pi}\right) 
  \ \ \ .
\label{eq:deutmag}
\end{eqnarray}
Comparision with the measured value of $\mu_M$ gives
\begin{eqnarray}
  \Ltwo (m_\pi) & = & -0.149\ {\rm fm^4}
\ \ \ \ ,
\end{eqnarray}
at the renormalization scale $\mu=m_\pi$.
The evolution of the $\Ltwo (\mu)$ operator as the renormalization scale is
changed is determined by the RG equation
\begin{eqnarray}
    \mu {d\over d\mu}
    \left[ { \Ltwo \over \left(\Czerominus\right)^2} \right] 
     & = & 0
\ \ \ \ .
\label{eq:L2RG}
\end{eqnarray}

The magnetic form factor has an expansion in powers of $Q$,
$F_M (q^2) = F_M^{(0)} (q^2) + F_M^{(1)} (q^2) +~...$.
We find that
\begin{eqnarray}
{e F_M^{(0)}(q^2)\over 2 M_d} & = &
\kappa_0 \ {e\over M_N}\ F_C^{(0)}({q}^2)\ =\
\kappa_0 \ {e\over M_N}
 {4\gamma\over\sqrt{-q^2}} \tan^{-1}\left(
    {\sqrt{-q^2}\over 4\gamma}\right)
\ \ \ \ ,
\end{eqnarray}
and 
\begin{eqnarray}
{e F_M^{(1)}(q^2)\over 2 M_d} & = &
\kappa_0 \ {e\over M_N} \  F_C^{(1)}({q}^2) 
\ + \ e \ \Ltwo \ {\gamma\over \pi} (\mu -\gamma)^2
\nonumber\\
& = & 
- \kappa_0 \ {e\over M_N} \ 
\gamma\rho_d \left[ 1 - 
{4\gamma\over\sqrt{-q^2}} \tan^{-1}\left( {\sqrt{-q^2}\over 4\gamma}\right)
\right]
\ +\
e \ \Ltwo \ {\gamma\over \pi} (\mu -\gamma)^2
\ \ \ \ .
\label{eq:magFF}
\end{eqnarray}

Despite the fact that $\Ltwo$ is numerically small, it is important to realize
that the effective range expansion for the deuteron magnetic moment
is not formally valid beyond leading order
(where  $\mu_M =  {e\over 2 M_N} \left(\kappa_p\ +\ \kappa_n\right)$).
At NLO, there is a counterterm that is allowed by the symmetries of
the theory and it is expected to be of natural size, which
in $\nopi$ is set by the pion mass.
Given that there is a counterterm at NLO, we do not pursue this calculation to
higher orders, even though it is straightforward.

%
\begin{figure}[t]
\centerline{{\epsfxsize=4.5in \epsfbox{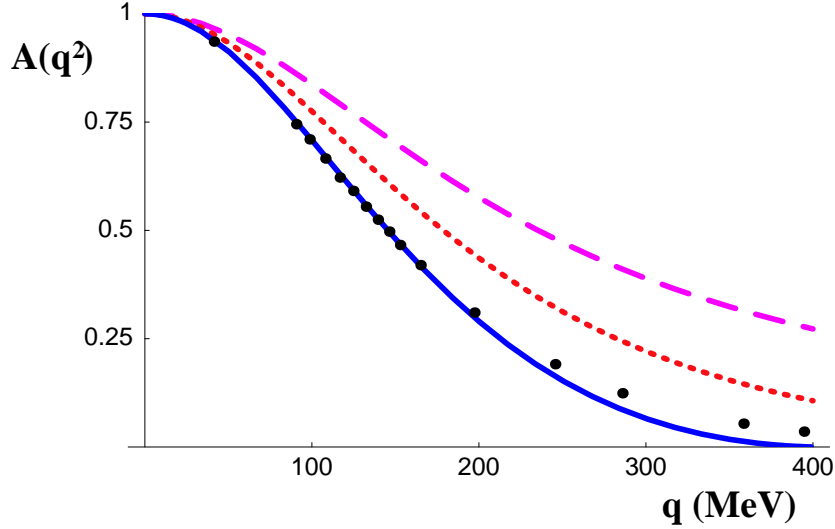}} }
\noindent
\caption{\it
The form factor $A( q^2)$ as  a function of 
$|{\bf q}|=\sqrt{-q^2}$.  
The dashed curve corresponds to the leading order prediction,
the dotted curve corresponds to the next-to-leading order prediction,
and the solid curve corresponds to the next-to-next-to-leading order 
prediction, in $\nopi$.
}
\label{fig:Aplot}
\vskip .2in
\end{figure}
%
%
\begin{figure}[t]
\centerline{{\epsfxsize=4.5in \epsfbox{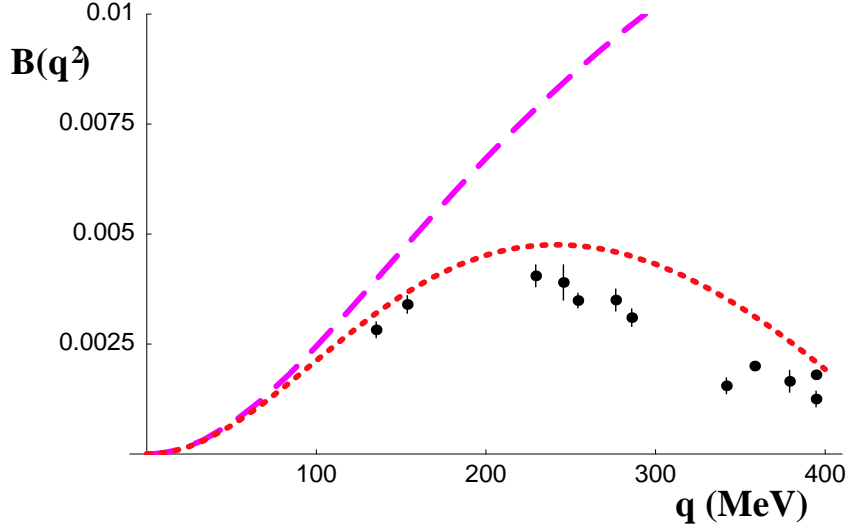}} }
\noindent
\caption{\it The form factor $B( q^2)$ as  a function of 
$|{\bf q}|=\sqrt{-q^2}$.  
The dashed curve corresponds to the leading order prediction,
 and the dotted curve corresponds to the next-to-leading order prediction, 
 in $\nopi$.
}
\label{fig:Bplot}
\vskip .2in
\end{figure}

It is combinations of the electric, magnetic and quadrupole form factors that
are measured in elastic electron-deuteron scattering.
The differential cross section for unpolarized
elastic electron-deuteron scattering is
given  by
\begin{eqnarray}
{{\rm d}\sigma\over {\rm d}\Omega} = {{\rm
d}\sigma\over {\rm d}\Omega}\biggl\vert_{\rm Mott}\biggr.\[A(q^2) +
B(q^2)\tan^2\left({\theta\over 2}\right)
\]
 \ \ \ \ ,
\label{eq:eddiff}
\end{eqnarray}
where $A$ and $B$ are related to the form factors
that appear in eq.~(\ref{eq:emmatdef})  by\cite{Wong}
\begin{eqnarray}
A & = & F_C^2 + {2\over 3}\eta  F_M^2 + {8\over 9}\eta^2 F_{\cal Q}^2
\ \ \ \  ,
\nonumber\\
B & = & {4\over 3}\eta(1+\eta) F_M^2\ ,
\label{eq:ABdef}
\end{eqnarray}
with $\eta = -q^2/4M_d^2$.  
In order to compare with data, we take our
analytic results for the form factors and expand the expression 
eq.~(\ref{eq:ABdef}) in powers of $Q$.
At the order we are working, 
$A$ is sensitive both the electric and magnetic
form factors, 
while $B$ depends only on the magnetic form factor.
The predictions for $A(q^2)$ and $B(q^2)$ along with data are shown
in fig.~(\ref{fig:Aplot}) and fig.~(\ref{fig:Bplot}), respectively.


\subsection{Electric Quadrupole Form-Factor of the Deuteron}

The quadrupole form factor is dominated by mixing between the
S-wave and D-wave components of the deuteron due to the NN interaction.
However, at subleading orders there are  contributions from two-nucleon-one
photon operators that do not contribute to nucleon-nucleon scattering (the same
as for the deuteron magnetic moment).
In the theory with pions, the leading contribution to the quadrupole form
factor is from the exchange of a potential pion\cite{KSW2}, occurring at NLO in
the power counting.
The two-loop graphs give
\begin{eqnarray}
{F_{\cal Q}^{(-1)} ({ |{\bf k}|^2 }) \over M_d^2} & = & 
{3 g_A^2 M_N\gamma \over  16 \pi f^2 {|{\bf k}|}^3}
\int_0^1 dx\
{1\over x \beta^4 \Delta }
\nonumber\\
& &
\times\left(
\left[ 3 {|{\bf k}|}^2 x^2 (1+\beta^2 )^2 - 24 {|{\bf k}|} m_\pi \beta   x (1+\beta^2 )
+ 16 m_\pi^2 \beta  ^2 (3+\beta^2 )\right] \tan^{-1} \beta
\right.
\nonumber\\
& & \left.
\ +\beta   \left[ -48 m_\pi^2\beta^2 + 8 m_\pi {|{\bf k}|} x \beta    (3+2\beta^2 )
- {|{\bf k}|}^2 x^2 (3+5\beta^2 )
\right]
\right)
\label{eq:formsub}
\end{eqnarray}
where
\begin{eqnarray}
\Delta(x)= \sqrt{\gamma^2\  +\ {1\over 4} x(1-x)|{\bf k}|^2}\ ,\qquad
\beta(x)    =  { {|{\bf k}|} x \over 2 (\gamma + m_\pi + \Delta  )}
\ \ \ \ .
\label{eq:deltadef}
\end{eqnarray}
One finds the quadrupole moment at this order to be 
\begin{eqnarray}
\mu_{\cal Q}^{(-1)} =
{g_A^2 M_N (6\gamma^2 + 9 \gamma m_\pi + 4 m_\pi^2)\over 30\pi f^2 (m_\pi+2
\gamma)^3}
\ \ \ \  .
\label{eq:quadmoment}
\end{eqnarray}
Numerically, this is
$\mu_{\cal Q} = \mu_{\cal Q}^{(-1)} =+0.40~{\rm fm^2}$,
about $30\%$ larger than
the measured value.  This magnitude of deviation is expected based on the size
of the expansion parameter.
Subleading contributions, at NNLO in the theory with pions, have been computed in
\cite{Binger}.  They depend upon the value of a coefficient that has not yet
been determined from nucleon-nucleon scattering in the theory with pions,
and so a numerical prediction at this order does not presently exist.
Notice, that the expression in eq.~(\ref{eq:quadmoment}) is of order $Q^{-1}$ 
in the theory with dynamical pions,
but is of order $Q^0$ in $\nopi$.

In $\nopi$, the contribution from single and multi-pion exchange and from all
other meson exchanges to the mixing between the S-wave and D-wave are captured
in the $\CSDzero$ and $\CSDtwotwo$ operators, up to NLO.
However, at NLO there is also a contribution from a
four-nucleon-one-photon operator, described by the Lagrange density
\begin{eqnarray}
  {\cal L}_{\cal Q} & = & -e\ \CQuad  (N^T P_i N)^\dagger (N^T P_j N)
\left(\nabla^i\nabla^j - {1\over n-1 }\nabla^2\delta^{ij}\right) A_0
\ \ \ .
\label{eq:quadct}
\end{eqnarray}
The operator connects two nucleons in initial and final S-wave states to a
D-wave photon.

As we have done with the other form factors, we expand the quadrupole form
factor in powers of $Q$, $ F_{\cal Q} =
F_{\cal Q}^{(0)} + F_{\cal Q}^{(1)} +
...$.
The leading order form factor is found to be 
\begin{eqnarray}
 {1\over M_d^2}  F_{\cal Q}^{(0)} ( |{\bf k}| ) & = &
-\CSDzeroone { M_N (\mu-\gamma)\over 32\pi |{\bf k}|^2}
\left[ -16\gamma^2 + (3  |{\bf k}|^2 + 16 \gamma^2) 
 {4\gamma\over  |{\bf  k}|}\tan^{-1}
   \left( { |{\bf k}|\over 4\gamma}\right)
   \right]
\ \ \ \ .
\label{eq:quadFFLO}
\end{eqnarray}
At subleading order, we  find a contribution to  the quadrupole form factor of
\begin{eqnarray}
  {1\over M_d^2}  F_{\cal Q}^{(1)} ( |{\bf k}|) & = &
  - \left[
    { \gamma(\mu-\gamma)^2\over \pi}\ \CQuad
    \right.\nonumber\\
& & \left.
  \ +\
    {M_N\gamma\rho_d (\mu-\gamma)\over 4\pi  |{\bf k}|^2}\    \CSDzeroone
   \left( -\gamma^2 +
    \left({3\over 16} |{\bf k}|^2 + \gamma^2 \right)
       {4\gamma\over  |{\bf  k}|}\tan^{-1}
   \left( { |{\bf k}|\over 4\gamma}\right)\ 
   \right)\ \right]
\ \ \ ,
\label{eq:QuadFF}
\end{eqnarray}
where we have written the form factor in terms of the three-momentum transfer
and not the four-momentum transfer as we do not encounter relativistic effects
at this order.

At LO and NLO the deuteron quadrupole moment, $\mu_{\cal Q}$, is found to be
\begin{eqnarray}
\mu_{\cal Q}^{(0)} & = &  \ -{\displaystyle{M_{N}\hspace{0.05cm}
\CSDzeroone (\mu -\gamma ) \over 12\pi }}
\nonumber\\
\mu_{\cal Q}^{(1)} & = &  
{1\over 2} \rho_d \gamma \ \mu_{\cal Q}^{(0)}
\ -\ \CQuad \ {(\mu-\gamma)^2\gamma\over\pi}
\quad .
\label{eq:Quadana}
\end{eqnarray}
Numerically, one finds that the quadrupole moment is given by
the sum of $ \mu_{\cal Q}^{(0)}  =  0.273\ {\rm fm^2}$ and
$ \mu_{\cal Q}^{(1)}  =  \left( 0.056 \  -  \ 0.0165\ \CQuad \right) \ {\rm fm^2}$
giving, at NLO, 
\begin{eqnarray}
\mu_{\cal Q} & = & \left( 0.329\ -\ 0.0165\ \CQuad \right) \ {\rm fm^2}
\ \ \ ,
\end{eqnarray}
where $\CQuad$ is measured in ${\rm fm^5}$.
Setting
$\CQuad=0$, we find a quadrupole moment of $0.329\ {\rm fm^2}$,
which is to be compared with the measured value of
$0.286\ {\rm fm^2}$.
In order to reproduce the measured  value of the quadrupole moment
$\CQuad =+2.60\ {\rm fm^5}$, at $\mu=m_\pi$.

%
\begin{figure}[t]
\centerline{{\epsfxsize=4.5in \epsfbox{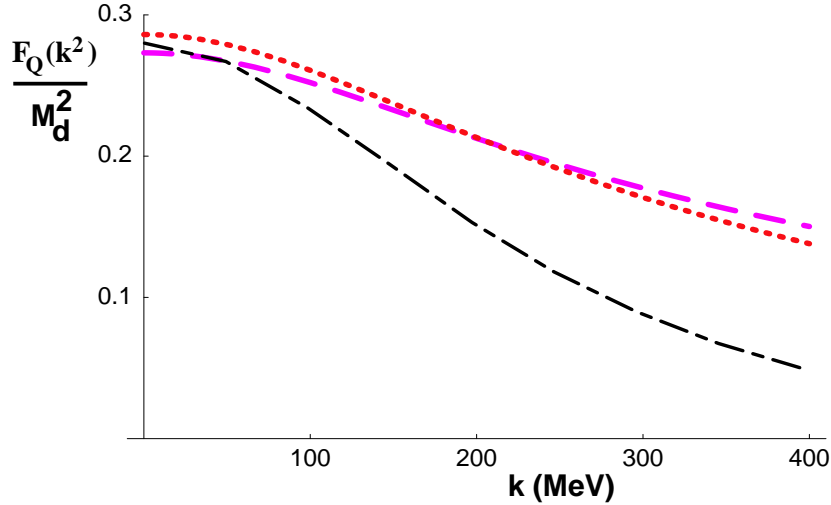}} }
\noindent
\caption{\it The quadrupole form factor $F_{\cal Q}( |{\bf k}|^2)$ 
as  a function of  $|{\bf k}|$.  
The dashed curve corresponds to the leading order prediction,
and the dotted curve corresponds to the next-to-leading order prediction, 
in $\nopi$.
The dot-dashed curve corresponds to a calculation with the Bonn-B potential
in the formulation of  \protect\cite{AGA} .
}
\label{fig:Qplot}
\vskip .2in
\end{figure}
In contrast to the electric and magnetic form factors, the quadrupole
form factor is found to deviate significantly from calculations
in potential models (e.g. the model of \cite{AGA})
for $|{\bf k}|\gsim m_\pi$, as shown  in fig.~(\ref{fig:Qplot}).
However, within the region $|{\bf k}|\lsim m_\pi$ where the $\nopi$ is
expected to perturbatively converge to the actual form factor
we see deviations of less than $\sim 10\%$, consistent with the size
of the expansion parameter.
It is clear that a NNLO calculation of this object is required in 
$\nopi$.


\subsection{The Radiative Capture Process $np\rightarrow d\gamma$ }

The radiative capture process $np\rightarrow d\gamma$ is a classic nuclear
physics demonstration of the existence of meson exchange currents.
In addition to the nucleon magnetic moment interactions which dominate this
cross section, photons minimally coupled to pions have been estimated to
contribute at the $\sim 10\%$  level in various schemes.
The cross section for this process was computed in the EFT with KSW power
counting in \cite{SSWst}, where it was found that the pion exchange current
contributions are
ultra-violet divergent and require the presence of a counterterm
at NLO (it would have been present at NLO even if the graphs were convergent).
In $\nopi$, there are no dynamical mesons, and hence no meson exchange
currents.
For this process, and all the observables discussed in this paper,
the effects of meson
exchanges are reproduced order by order by local operators involving
multiple nucleon fields.  At NLO in $\nopi$ this corresponds to a single
insertion of the $\Lone$ operator, defined in eq.~(\ref{eq:Ldef}).

The amplitude for
the radiative capture of extremely low momentum neutrons
$np\rightarrow d\gamma$
has contributions from both the $\si$ and $\siii$ $NN$
 channels. It can be written as
\begin{eqnarray}
i{\cal A}(np\rightarrow d\gamma) & = &
e\ X\ N^T\tau_2\ \sigma_2 \  \left[ {\bbox \sigma}\cdot {\bf k}\ 
\ {\bbox\epsilon} (d)^* \cdot {\bbox \epsilon} (\gamma)^*
  \ -\ {\bbox \sigma} \cdot  {\bbox \epsilon} (\gamma)^*\ 
  \ {\bf k}\cdot {\bbox \epsilon} (d)^* 
  \right] N 
\\ \nonumber
& + &
i e\ Y\  \epsilon^{ijk}\ \epsilon (d)^{i*}\   
k^j\  {\bbox\epsilon} (\gamma)^{k*}
\ (N^T\tau_2 \tau_3 \sigma_2 N)
\ \ \ \ ,
\label{eq:matrix}
\end{eqnarray}
where  $e=|e|$ is the magnitude of the  electron charge, 
$N$ is 
the doublet of nucleon spinors, ${\bbox \epsilon}(\gamma)$ is 
the polarization vector for the photon, ${\bbox \epsilon}(d)$
is the polarization
vector for the deuteron and ${\bf k}$ is the outgoing photon momentum.
The term with coefficient $X$ corresponds to capture from the $\siii$ channel
while the term with coefficient $Y$ corresponds to capture from the $\si$
channel.
For convenience, we define dimensionless variables $\tilde X$ and $\tilde Y$,
by
\begin{eqnarray}
  X & = & i {2\over M_N} \sqrt{\pi\over\gamma^3}\ \tilde X
  \ \ ,\ \ 
  Y =  i {2\over M_N} \sqrt{\pi\over\gamma^3}\ \tilde Y
  \ \ \ \ .
  \label{eq:XYdef}
\end{eqnarray}
Both $\tilde X$ and $\tilde Y$ have the $Q$ expansions,
$\tilde X = \tilde X^{(0)}+ \tilde X^{(1)} +...$,
and $\tilde Y=\tilde Y^{(0)}+ \tilde Y^{(1)}+...$,
where a superscript denotes the order in the $Q$ expansion.
The capture cross section for very low momentum
neutrons with speed $|{\bf v}|$ 
arising from eq.~(\ref{eq:matrix}) is
\begin{eqnarray}
  \sigma & = & {8\pi \alpha \gamma^3 \over M_N^5 |{\bf v}|}
  \left[ 2 |\tilde X|^2\ +\ |\tilde  Y|^2\right]
  \ \ \ ,
\label{eq:sig}
\end{eqnarray}
where $\alpha$ is the fine-structure constant.

%
%
\begin{figure}[t]
\centerline{{\epsfxsize=4.0in \epsfbox{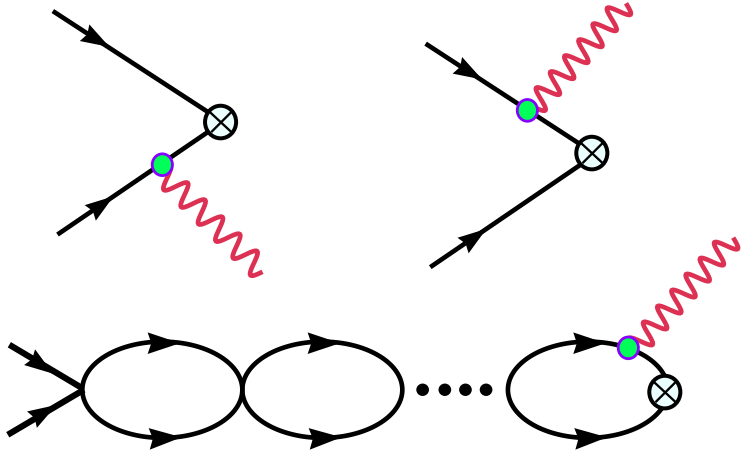}} }
\noindent
\caption{\it The Feynman diagrams giving the leading order contribution
to $np\rightarrow d\gamma$ in $\nopi$.
The solid lines denote nucleons and the wavy lines denote photons.
The light solid circles correspond to the nucleon magnetic
moment coupling of the photon.
The crossed circle represents an insertion of the deuteron
  interpolating 
  field . 
}
\label{fig:nplead}
\vskip .2in
\end{figure}
At leading order in $\nopi$ the amplitudes receive contributions from the 
Feynman diagrams shown in fig.~(\ref{fig:nplead}) and  are 
\begin{eqnarray}
  \tilde Y^{(0)} & = & \kappa_1\ \left( 1 - \gamma a^{(\si)}\right)
\ \ \ , \ \ \
 \tilde X^{(0)}\ =\ 0
\ \ \ ,
\label{eq:Ylead}
\end{eqnarray}
where $a^{(\si)} = -23.714\pm 0.013~{\rm fm}$, is the
scattering length in the
$\si$ channel, and $\kappa_1$ is the isovector magnetic moment defined in
eq.~(\ref{eq:Nucmag}).
\begin{figure}[t]
\centerline{{\epsfxsize=3.0in \epsfbox{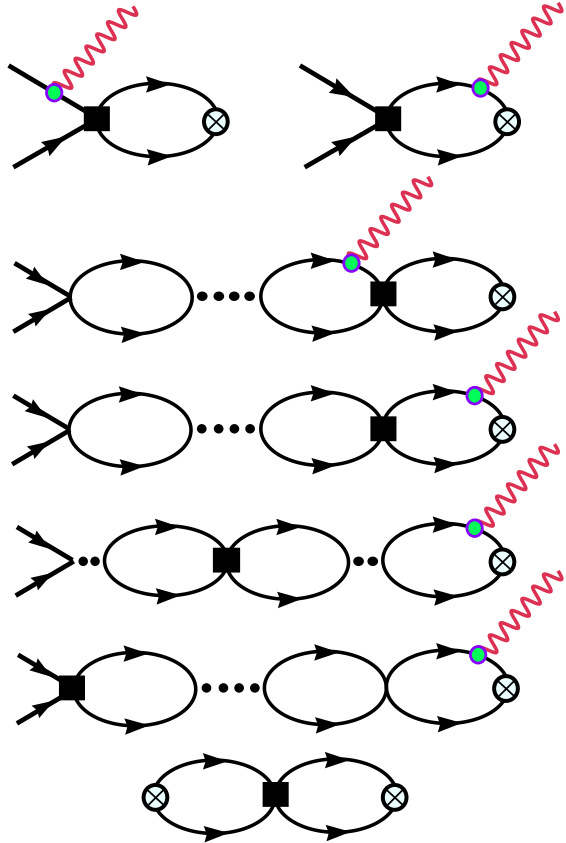}} }
\noindent
\caption{\it Graphs contributing to the amplitude  
  for $n+p\rightarrow d+\gamma$ at subleading order 
  due to insertions of the $C_2$ operators.
  The solid lines denote nucleons
  and the wavy lines denote photons.
The light solid circles correspond to the nucleon magnetic
moment coupling of the photon.
The solid square denotes a $C_2$ operator. 
The crossed circle represents an insertion of the deuteron
  interpolating 
  field . The last graph denotes the contribution from wavefunction
renormalization. 
  }
\label{fig:strongsubc2}
\vskip .2in
\end{figure}
%
\begin{figure}[t]
\centerline{{\epsfxsize=3.5in \epsfbox{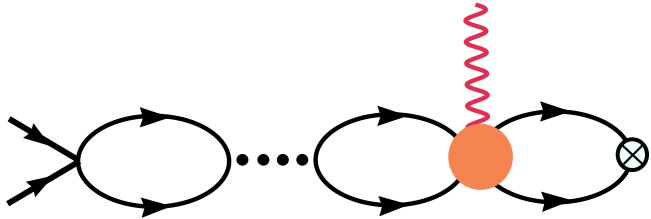}} }
\noindent
\caption{\it Local counterterm contribution to the amplitude  
  for $n+p\rightarrow d+\gamma$ at NLO.
  The solid lines denote nucleons
  and the wavy lines denote photons.
The solid circle corresponds to an insertion of the $\Lone$ operator.
  The crossed circle represents an insertion of the deuteron
  interpolating 
  field. }
\label{fig:strongsubL1}
\vskip .2in
\end{figure}
At next-to-leading order, NLO, the contribution 
arising from the Feynman diagrams shown in 
fig.~(\ref{fig:strongsubc2}) and fig.~(\ref{fig:strongsubL1})
is found  to
be\cite{SSWst}
\begin{eqnarray}
  \tilde Y^{(1)} & = & {1\over 2} \kappa_1\gamma\rho_d \left( 1 - \gamma
    a^{(\si)}\right)
  \nonumber\\
& - &  {M_N\over 4\pi} \gamma^2 a^{(\si)} \left(\mu-\gamma\right)
\left(\mu- {1\over a^{(\si)}}\right)
\left[ \Lone\ -\ {\kappa_1 \pi\over M_N} \left( { r_0^{(\si)}\over  (\mu- {1\over
        a^{(\si)}})^2}
    \ +\ {\rho_d\over  (\mu-\gamma)^2}\right)\right]
\ \ \ ,
\label{eq:YNLO}
\end{eqnarray}
where $r_0^{(\si)}$ is the effective range in the $\si$ channel.
We have not computed  $\tilde X^{(1)}$ as it can only contribute at NNLO since
$\tilde X^{(0)}$ vanishes.
The RG evolution of  $\Lone$ was discussed at length in \cite{SSWst}, where it
was made clear that its behavior is much different from  $\Ltwo$, the counterterm
for the deuteron magnetic moment.
In the absence of pions we find
\begin{eqnarray}
    \mu {d\over d\mu}
  \left[  { \Lone  - {1\over 2}\kappa_1 \left( \Ctwozero \ +\ 
       \Ctwozerosing  \right)\over
        \Czerominussing\ \Czerominus  }\right]
& = & 0
\ \ \ ,
\label{eq:LoneRG}
\end{eqnarray}
in order that the cross section for  $NN\rightarrow NN\gamma$
with the initial
nucleons in the $\si$ channel and the final nucleons in the  $\siii$ channel be
independent of the renormalization scale at all energies.
The analytic structure of the amplitude ensures that the capture cross section
will be  $\mu$-independent, if  $NN\rightarrow NN\gamma$
is  $\mu$-independent.

The cross section for this process has been measured very precisely
for an incident neutron speed of  $|{\bf v}| = 2200\ {\rm m/s}$
to be $\sigma^{\rm expt} = 334.2\pm 0.5\ {\rm mb}$\cite{CWCa}. 
In $\nopi$ we find a cross section at NLO,
at this incident neutron speed, of
\begin{eqnarray}
\sigma_{\pislashsmall} & = & \left( 287.1\ +\ 6.51\ \Lone\ \right)\ {\rm mb}
\ \ \ \ ,
\label{eq:npsig}
\end{eqnarray}
where  $\Lone$ is in units of ${\rm fm}^4$ and is renormalized $\mu=m_\pi$. 
Requiring  $\sigma_{\pislashsmall}$ to reproduce the measured cross section 
$\sigma^{\rm expt}$ fixes $\Lone = 7.24~{\rm fm}^4$.

We see that even in the theory without dynamical pions, one is able to recover
the cross section for radiative neutron capture at higher orders.
It is clear that in this theory the four-nucleon-one-photon operators play a
central role in reproducing the low energy observables.
In the theory with pions, one can see by examining the contributing
Feynman diagrams\cite{SSWst},
that in the limit that the momentum transferred to the photon is small the
pion propagators can be replaced by $1/m_\pi^2$, while keeping the derivative
structure in the numerator.
This contribution, as well as the contribution from all hadronic exchanges,
is reproduced order by order in
the momentum expansion by the contributions from local multi-nucleon-photon
interactions.
From the calculations in the theory with dynamical pions, the value of $\Lone$
is not saturated by pion exchange currents as these contributions are
divergent, and require the presense of the $L_1$ operator\cite{SSWst}.
Therefore, estimates of $\Lone$ based on meson exchanges alone are
model dependent.

The effective range calculation of $np\rightarrow d\gamma$ was first performed
by Bethe and Longmire\cite{ERtheory} and revisited by Noyes\cite{Noyes}.
After correcting the typographical errors in the
expression for $\sigma$ that appears in
the Noyes article, the expressions in the two papers\cite{ERtheory,Noyes}
are identical,
\begin{eqnarray}
  \sigma^{( {\rm ER})} & = &
  {2\pi\alpha  \ \kappa_1^2 \ \gamma^6\ (a^{(\si)})^2\  a^{(\siii)}
    \over |{\bf v}|
    M_N^5 \left( 2\ -\ \gamma a^{(\siii)}\right)  }
  \left( 1\ +\ {1\over \gamma a^{(\siii)}}\ -\ {2\over\gamma a^{(\si)}} -
    {1\over 2}\gamma r_0^{(\si)} \right)^2
\ \ \ ,
\label{eq:sigER}
\end{eqnarray}
which when expanded in powers of $Q$ is
\begin{eqnarray}
  \sigma^{( {\rm ER})} & = &
{8\pi\alpha \gamma^3\over|{\bf v}|  M_N^5 }
\left[
  \kappa_1^2  (1-\gamma a^{(\si)})^2
\right.\nonumber\\
& & \left.
  \ +\
  {1\over 2}\gamma \left(\rho_d-r_0^{(\si)}\right) \kappa_1^2 (1-\gamma a^{(\si)})^2
  \ +\
  {1\over 2}\gamma \left(\rho_d+r_0^{(\si)}\right)\kappa_1^2 (1-\gamma
  a^{(\si)})
  \ +\ ...\ 
\right]
\ \ \ .
\label{eq:sigERexp}
\end{eqnarray}
From this expansion, one finds that
\begin{eqnarray}
  \tilde Y^{{\rm ER} , (0)} & = & \kappa_1\ \left( 1 - \gamma a^{(\si)}\right)
\ \ \ , \ \ \
 \tilde X^{{\rm ER} ,(0)}\ =\ 0
\ \ \ ,
\label{eq:YleadER}
\end{eqnarray}
and
\begin{eqnarray}
  \tilde Y^{{\rm ER} ,(1)} & = &
  {1\over 4} \kappa_1\gamma \left( \rho_d-r_0^{(\si)}\right)
    \left( 1 - \gamma a^{(\si)}\right)
\ +\ 
  {1\over 4} \kappa_1\gamma \left( \rho_d+ r_0^{(\si)} \right)
\ \ \ .
\label{eq:YNLOER}
\end{eqnarray}

At LO in the $\nopi$ expansion the cross sections agree, however, at
NLO the expressions are very different.
In addition to the counterterm that appears at this order in the $\nopi$,
the contributions
from the effective range parameters are found to disagree.
In the $\nopi$ the local counterterm is renormalized by the short-distance
behavior of graphs involving the $C_2$ operators and hence the effective range
parameters in both channels.  Given, this behavior it is no surprise that the
effective range contributions differ between the two calculations.
Given the explicit $\mu$-dependence in $\tilde Y^{(1)}$ and the absence of such
dependence in
$ \tilde Y^{{\rm ER} ,(1)}$ it is amusing to ask if there is a scale for
which
the expressions are identical, with $\Lone=0$.
Indeed such a scale exits,
\begin{eqnarray}
\mu^{{\rm ER}} & = &  {  \gamma  r_0^{(\si)}  a^{(\si)} - \rho_d\over
   a^{(\si)} \left( r_0^{(\si)} -  \rho_d\right)}
\ \ \  ,
\label{eq:musame}
\end{eqnarray}
which, by inserting the appropriate values, gives  scale $\mu^{{\rm ER}}\sim
144~{\rm MeV}$, coincidentally close to $\mu=m_\pi$.


\section{Conclusions}

An effective field theory without dynamical pions can describe all
low-energy two-nucleon processes, and generally all multi-nucleon processes.
We have shown in detail how $\nopi$ systematically reproduces nucleon-nucleon
scattering and the interactions of the deuteron.
Effective range theory is seen to emerge as the uncontrolled approximation to
$\nopi$ where the multi-nucleon-external current local operators are neglected.
For most observables, this is a small effect, while for the radiative capture
process $np\rightarrow d\gamma$ this omission gives rise to 
deviations at the $10\%$ level, which
has been known for decades.
Inclusion of the local four-nucleon-one-magnetic-photon interaction which
enters at NLO in $\nopi$ introduces a free parameter $\Lone$, which can be fit
to reproduce the measured cross section.
As a consequence of this ultra-violet behavior, the NLO contribution to the
cross section in $\nopi$ with $\Lone=0$
is unrelated to that determined by effective range theory with the simple
neglect of such contributions.
While this does not satisfy our desire to make a parameter free prediction for
this cross section beyond leading order,
it does allow us to rigorously relate S-matrix elements for different processes
based solely on the symmetries of QCD.
The deuteron magnetic moment is another example of  an observable where a
counter term $\Ltwo$ appears at NLO, but numerically this is found to be small.

In working to higher orders in $\nopi$ one encounters relativistic effects,
starting at NNLO.
The contribution of relativistic effects on nucleon-nucleon scattering, the
electric charge form factor of the deuteron and the electric polarizability of
the deuteron were calculated at leading order.
Their impact upon the static properties of the deuteron are expected to be
very small, of order $\sim\gamma^2/M_N^2$, and this was recovered.

We should take a moment to ask about the impact this work will have on our
understanding of potential model approaches to nucleon interactions.
As we have emphasized, for most processes the effective range expansion
provides a relatively good approximation.
For some of
the observables, such as the deuteron charge radius and the electric
polarizability, this arises because counterterms only appear at very high orders
in the momentum expansion, while for others, such as the
$np\rightarrow d\gamma$, and the deuteron quadrupole moment,
there is a four-nucleon-one-photon
counterterm that prevents effective range theory from doing
better than $\sim 10\%$.
Such counterterms cannot be fixed by the nucleon-nucleon scattering amplitude no
matter how precisely it is measured, but they can be determined from
inelastic scattering processes.
Thus, multi-nucleon 
processes involving external currents receive contributions
from interactions beyond those related by gauge invariance to the 
nucleon-nucleon interaction.
The form and magnitude of such interactions will, in general,  depend
explicitly upon the
choice of the nucleon-nucleon interaction, as is made clear by the RG behavior
of the $\Lone$ counterterm.

\vskip 0.5in

We thank D. Phillips for many useful discussions and for providing 
us with the calculation of the quadrupole form factor in the formulation
of \cite{AGA} using the Bonn-B potential, that appears in 
fig.~(\ref{fig:Qplot}).
We would like to thank David Kaplan and Mark Wise for 
several interesting discussions. 
This work is supported in part by the U.S. Dept. of Energy under
Grants No. DE-FG03-97ER4014.

\end{document}